\documentclass[apj]{emulateapj}
\newcommand{\kms}{\ifmmode{~{\rm km~s^{-1}}}\else{~km s$^{-1}$}\fi}
\newcommand{\cubecm}{\ifmmode{~{\rm cm^{-3}}}\else{~cm$^{-3}$}\fi}
\newcommand{\lsim}{\lower0.3em\hbox{$\,\buildrel <\over\sim\,$}}
\newcommand{\gsim}{\lower0.3em\hbox{$\,\buildrel >\over\sim\,$}}

\newcommand{\flux}{erg s$^{-1}$ cm$^{-2}$ Hz$^{-1}$}

\newcommand{\enzo}{{\sl Enzo}}
\newcommand{\Ms}{\ifmmode{M_\odot}\else{$M_\odot$}\fi}

\newcommand{\hh}{H$_2$}
\newcommand{\Ol}{$\Omega_\Lambda$}
\newcommand{\Om}{$\Omega_M$}
\newcommand{\Ob}{$\Omega_b$}

\newcommand{\tcool}{$t_{\rm{cool}}$}

\newcommand{\tdyn}{$t_{\rm{dyn}}$}

\newcommand{\tvir}{$T_{\rm{vir}}$}
\newcommand{\rvir}{\ifmmode{{\rm r_{vir}}}\else{r$_{\rm{vir}}$}\fi}
\newcommand{\mvir}{\ifmmode{{\rm M_{vir}}}\else{M$_{\rm{vir}}$}\fi}

\newcommand{\lya}{Ly$\alpha$}
\newcommand{\jj}{\ifmmode{J_{21}}\else{$J_{21}$}\fi}
\newcommand{\flw}{\ifmmode{F_{\rm{LW}}}\else{$F_{\rm{LW}}$}\fi}
\newcommand\zstar[1]{$10^{#1}f_{80}$}
\newcommand\tento[1]{$10^{#1}$}
\newcommand{\feighty}{\ifmmode{f_{80}}\else{$f_{80}$}\fi}

\begin{document}

\shorttitle{RESOLVING THE FORMATION OF PROTOGALAXIES III}
\shortauthors{WISE \& ABEL}

\title{Resolving the Formation of Protogalaxies. III. Feedback from
  the First Stars}
\author{John H. Wise\altaffilmark{1,2} and Tom Abel\altaffilmark{1}}

\altaffiltext{1}{Kavli Institute for Particle Astrophysics and
  Cosmology, Stanford University, Menlo Park, CA 94025}
\altaffiltext{2}{Laboratory for Observational Cosmology, NASA Goddard
  Space Flight Center, Greenbelt, MD 21114}
\email{john.h.wise@nasa.gov}

\begin{abstract}

  The first stars form in dark matter halos of masses $\sim$$10^6 \Ms$
  as suggested by an increasing number of numerical simulations.
  Radiation feedback from these stars expels most of the gas from
  their shallow potential well of their surrounding dark matter halos.
  We use cosmological adaptive mesh refinement simulations that
  include self-consistent Population III star formation and feedback
  to examine the properties of assembling early dwarf galaxies.
  Accurate radiative transport is modelled with adaptive ray tracing.
  We include supernova explosions and follow the metal enrichment of
  the intergalactic medium.  The calculations focus on the formation
  of several dwarf galaxies and their progenitors.  In these halos,
  baryon fractions in 10$^8$ \Ms~halos decrease by a factor of 2 with
  stellar feedback and by a factor of 3 with supernova explosions.  We
  find that radiation feedback and supernova explosions increase
  gaseous spin parameters up to a factor of 4 and vary with time.
  Stellar feedback, supernova explosions, and \hh~cooling create a
  complex, multi-phase interstellar medium whose densities and
  temperatures can span up to 6 orders of magnitude at a given radius.
  The pair-instability supernovae of Population III stars alone enrich
  the halos with virial temperatures of 10$^4$ K to approximately
  10$^{-3}$ of solar metallicity.  We find that 40\% of the heavy
  elements resides in the intergalactic medium (IGM) at the end of our
  calculations.  The highest metallicity gas exists in supernova
  remnants and very dilute regions of the IGM.

\end{abstract}

\keywords{cosmology: theory --- galaxies: dwarf --- galaxies:
  high-redshift --- stars: formation}

\section{MOTIVATION}

The majority of galaxies in the universe are low-luminosity, have
masses of $\sim$$10^8$ solar masses, and are known as dwarf galaxies
\citep{Schechter76, Ellis97, Mateo98}.  Galaxies form hierarchically
through numerous mergers of smaller counterparts \citep{Peebles68,
  White78}, whose properties will inevitably influence their parent
galaxy.  Dwarf galaxies are the smallest galactic building blocks, and
this leads to the question on even smaller scales: how were dwarf
galaxies influenced by their progenitors?

A subset of dwarf galaxies, dwarf spheroidals (dSph), have the highest
mass-to-light ratios \citep{deBlok97, Mateo98} and contain a
population of metal-poor stars that are similar to Galactic halo stars
\citep{Tolstoy04, Helmi06}.  There is a metallicity floor exists of
$10^{-3}$ and $10^{-4}$ of solar metallicity in dSph and halo stars,
respectively \citep{Beers05, Helmi06}.  Stellar metallicities increase
with time as previous stars continually enrich the interstellar medium
(ISM).  Hence the lowest metallicity stars are some of the oldest
stars in the system and can shed light on the initial formation of
dwarf galaxies.  This metallicity floor also suggests that metal
enrichment was widespread in dark matter halos before low-mass stars
could have formed \citep[e.g.][]{Ricotti02b}.  Supernovae (SNe) from
metal-free (Pop III) stars generate the first metals in the universe
and may supply the necessary metallicity to form the most metal-poor
stars observed \citep{Ferrara98, Madau01a, Norman04}.

Dwarf galaxy formation can be further constrained with observations
that probe reionization and semi-analytic models.  Observations of
luminous quasars powered by supermassive black holes (SMBH) of mass
$\sim$$10^9 \Ms$ \citep{Becker01, Fan02, Fan06} and low-luminosity
galaxies \citep{Hu02, Iye06, Kashikawa06, Bouwens06, Stark07} at and
above redshift 6 indicate that active star and BH formation began long
before this epoch.  Semi-analytic models have argued that cosmological
reionization was largely caused by low-luminosity dwarf galaxies
\citep{Haiman97, Cen03, Somerville03, Wise05, Haiman06}.  Some of the
most relevant parameters in these models control star formation rates,
ionizing photon escape fractions, metal enrichment, and the minimum
mass of a star forming halos.  They are usually constrained using (i)
the cosmic microwave background (CMB) polarization observation from
Wilkinson Microwave Anisotropy Probe (WMAP) that measures the optical
depth of electron scattering to the CMB \citep{Page06}, (ii)
Gunn-Peterson troughs in $z \sim 6$ quasars, and (iii) numerical
simulations that examine negative and positive feedback of radiation
backgrounds \citep{Machacek01, Machacek03, Yoshida03, Mesigner06}.
Radiation hydrodynamical \textit{ab initio} simulations of the first
stars \citep{Yoshida06a, Abel07} and galaxies can further constrain
the parameters used in semi-analytic models by analyzing the impact of
stellar feedback on star formation rates and the propagation of
\ion{H}{2} regions in the early universe.  Moreover, these simulations
contain a wealth of information pertaining to the properties of Pop
III star forming halos and early dwarf galaxies that can increase our
understanding of the first stages of galaxy formation.

First we need to consider Pop III stars, which form in the progenitor
halos of the first galaxies, to capture the initial properties of
dwarf galaxies.  Cosmological numerical studies have shown that
massive (30--300 \Ms) Pop III stars form in dark matter halos with
masses $\sim$$10^6 \Ms$ \citep{Abel02, Bromm02, Yoshida06b, Gao07,
  OShea07}.  Recently, \citeauthor{Yoshida06b} and \citet{Turk08}
followed the gaseous collapse of a molecular cloud that will host a
Pop III star to cosmologically high number densities of $10^{16}$ and
$10^{21}$ \cubecm, respectively.  The former group thoroughly analyzed
the gas dynamics, cooling, and stability of this free-fall collapse.
The latter group observed a protostellar core forming with 10 Jupiter
masses that is bounded by a highly asymmetric protostellar shock.
Both groups found no fragmentation in the fully molecular core that
collapses into a single, massive $\sim$100\Ms~star.  Furthermore,
\citet{Omukai03} determined that accretion may halt at the same mass
scale, using protostellar models even for different mass accretion
histories.

Pop III stars with stellar masses roughly between 140 and 260 \Ms~end
their life in a pair-instability SN (PISN) that releases $10^{51} -
10^{53}$ ergs of energy and tens of solar masses of heavy elements
into the ambient medium \citep{Barkat67, Bond84, Heger02}.  These
explosions are an order of magnitude larger than typical Type II SNe
in both quantities \citep{Woosley86}, such explosions energies are
larger than the binding energies in their low-mass hosts, e.g., $2.8
\times 10^{50}$ ergs for a $10^6 \Ms$ halo at redshift 20.  Thus gas
structures in the host halo are totally disrupted and expelled,
effectively enriching the surrounding intergalactic medium (IGM) with
the SN ejecta \citep{Bromm03, Kitayama05, Greif07}.  The combination
of the shallow potential well and large explosion energy suggests that
these events are good candidates for enriching the first galaxies and
IGM.  Outside of the pair-instability mass range, Pop III stars die by
directly collapsing into a BH \citep{Heger03}, possibly providing the
seeds of high-redshift quasars in galaxies that are associated with
the rarest density fluctuations \citep[e.g.][]{Madau01b, Volonteri05,
  Trenti07}.

One-dimensional calculations \citep{Whalen04, Kitayama04, Kitayama05}
and recent three-dimensional radiation hydrodynamical simulations
\citep{Yoshida06a, Abel07} have investigated how the Pop III stellar
feedback affects its host halo and nearby cosmic structure.  In
addition to SNe, \ion{H}{2} regions surrounding Pop III stars, which
have luminosities $\sim$$10^6 L_\odot$ \citep{Schaerer02}, alone can
dynamically affect gas at distances up to a few proper kpc.
Ionization fronts and \ion{H}{2} regions \citep[see][for a
  review]{Yorke86} have been extensively studied in literature on star
formation since \citet{Stroemgren39}.  Stellar radiation generates an
ionization front that begins as a R-type front and transforms into a
D-type front when its speed slows to twice the sound speed of the
ionized gas.  Then a strong shock wave forms at the front and recedes
from the star at $\sim$$30\kms$.  The ionization front decouples from
the shock wave and creates a final \ion{H}{2} region that is 1 -- 3
proper kpc in radius for massive Pop III stars residing in low-mass
halos.  The ionized gas is warm ($\sim$$3 \times 10^4$ K) and diffuse
($\sim$$1 \cubecm$).  The shock wave continues to accumulate gas and
advance after the star dies.  Eventually it stalls in the IGM, but in
the process, it reduces the baryon fraction of the halo below one
percent \citep{Yoshida06a, Abel07}.

Clearly the number of progenitors of a given galaxy as well as the
star formation and feedback history of the progenitors will play a
role in shaping all of its properties.  But how much?  If most stars
of a galaxy are formed later, will the earliest episodes not be
entirely negligible?  To start addressing these questions, we have
carried out a suite of simulations that include accurate three
dimensional radiative transfer and the SN explosions of Pop III stars
and have followed the buildup of several dwarf galaxies from those Pop
III star hosting progenitors. The Pop III radiative and SN feedback
dramatically alter the properties of high redshift dwarf galaxies,
and we discuss some of the most striking differences here.  We leave a
more detailed exposition of star formation rates, star forming
environments, and the beginning of cosmic reionization for a later
paper.

% A few of these local dSph's (Tucana, Cetus, and KKR 25) reside far
% from the Milky Way and Andromeda galaxies.  They are thought to be
% undisturbed since its formation, truly being relics of the
% high-redshift universe.

In the following section, we detail our cosmological, radiation
hydrodynamics simulations and the star formation algorithm.  Then we
describe the global characteristics of dwarf galaxies that forms in
our simulations in \S\ref{sec:reionResults}.  There we also focus on
metal enrichment of star forming halos and the IGM, arising from
PISNe.  In \S\ref{sec:reionDiscuss}, we discuss the implications of
our findings on the paradigm of high-redshift galaxy formation by
including \hh~chemistry and Pop III star formation and feedback.  We
summarize in the last section.

\section{RADIATION HYDRODYNAMICAL SIMULATIONS}

We use the Eulerian AMR hydrodynamic code \enzo~\citep{Bryan97,
  Bryan99} to study the importance of primordial stellar feedback in
early galaxy formation.  \enzo~uses an $n$-body adaptive particle-mesh
solver \citep{Couchman91} to follow the dark matter (DM) dynamics.  We
first describe the setup of our simulations.  We then detail our star
formation recipe for primordial star formation.  Also we have
implemented adaptive ray tracing into \enzo~whose description
concludes this section.

%%%%%%%%%%%%%%%%%%%%%%%%%%%%%%%%%%%%%%%%%%%%%%%%%%%%%%%%%%%%%%%%%%%%%%%%
%
% SIMULATION DETAILS
%
%%%%%%%%%%%%%%%%%%%%%%%%%%%%%%%%%%%%%%%%%%%%%%%%%%%%%%%%%%%%%%%%%%%%%%%%

\begin{deluxetable*}{lcccccc}
%\tablecolumns{7}
\tabletypesize{}
\tablewidth{0pc}
\tablecaption{Simulation Parameters\label{tab:sims}}

\tablehead{
  \colhead{Name} & \colhead{$l$} & \colhead{SF} & \colhead{SNe}
  & \colhead{N$_{\rm{part}}$} & \colhead{N$_{\rm{grid}}$} & 
  \colhead{N$_{\rm{cell}}$} \\
  \colhead{} & \colhead{[Mpc]} & \colhead{} & \colhead{} & \colhead{}
  & \colhead{} & \colhead{} 
} 
\startdata

SimA-HHe & 1.0 & No & No & 2.22 $\times$ 10$^7$ & 40601 & 1.20
$\times$ 10$^8$ (494$^3$) \\
SimA-RT & 1.0 & Yes & No & 2.22 $\times$ 10$^7$ & 44664 & 1.19
$\times$ 10$^8$ (493$^3$) \\
SimB-HHe & 1.5 & No & No & 1.26 $\times$ 10$^7$ & 21409 & 6.51
$\times$ 10$^7$ (402$^3$) \\
SimB-RT & 1.5 & Yes & No & 1.26 $\times$ 10$^7$ & 24013 & 6.54
$\times$ 10$^7$ (403$^3$) \\
SimB-SNe & 1.5 & Yes & Yes & 1.26 $\times$ 10$^7$ & 24996 & 6.39
$\times$ 10$^7$ (400$^3$) 
\enddata
\tablecomments{Col. (1): Simulation name. Col. (2): Box
  size. Col. (3): Star formation. Col. (4): Supernova
  feedback. Col. (5): Number of dark matter particles. Col. (6):
  Number of AMR grids. Col. (7): Number of unique grid cells.}
\end{deluxetable*}
%%%%%%%%%%%%%%%%%%%%%%%%%%%%%%%%%%%%%%%%%%%%%%%%%%%%%%%%%%%%%%%%%%%%%%%%

\subsection{Simulation Setup}

We perform two cosmological realizations with different box sizes and
random phases and WMAP 1 year parameters of ($h$, \Ol, \Om, \Ob,
$\sigma_8$, $n$) = (0.72, 0.73, 0.27, 0.024$h^{-2}$, 0.9, 1)
\citep{Spergel03}.  Table \ref{tab:sims} summarizes the details of
these simulations.  The characteristics of the individual halos
studied here are not affected by the significantly different WMAP
third year parameters \citep[WMAP3;][]{Spergel06}, which do affect the
statistical properties of such halos.  We have verified that nothing
atypical occurs during the assembly of the halos studied here
\citep[see][hereafter Paper I]{Wise07a}.

The initial conditions are the same as in Paper I.  To simplify the
discussion, simulation A will always be quoted first with the value
from simulation B in parentheses.  They both have a top grid with a
resolution of 128$^3$ with three nested subgrids with twice finer
resolution and are initialized at z = 129 (119) with the COSMICS
package \citep{Bertschinger95, Bertschinger01}.  The box size is 1.0
(1.5) comoving Mpc.  The innermost grid has an effective resolution of
1024$^3$ with DM particle masses of 30 (101) \Ms and a side length of
250 (300) comoving kpc.

Regions of the simulation grid are refined by two when one or more of
the following conditions are met: (1) baryon density is greater than 3
times $\Omega_b \rho_0 N^{l(1+\phi)}$, (2) DM density is greater than
3 times $\Omega_{\rm{CDM}} \rho_0 N^{l(1+\phi)}$, and (3) the local
Jeans length is less than 16 cell widths.  Here $N = 2$ is the
refinement factor; $l$ is the AMR refinement level; $\phi = -0.3$
causes more frequent refinement with increasing AMR levels,
i.e. super-Lagrangian behavior; $\rho_0 = 3H_0^2/8\pi G$ is the
critical density; and the Jeans length, $L_J = \sqrt{15kT/4\pi\rho G
  \mu m_H}$, where $H_0$, $k$, T, $\rho$, $\mu$, and $m_H$ are the
Hubble constant, Boltzmann constant, temperature, gas density, mean
molecular weight in units of the proton mass, and hydrogen mass,
respectively.  The Jeans length refinement insures that we meet the
Truelove criterion, which requires the Jeans length to be resolved by
at least 4 cells on each axis \citep{Truelove97}.  Further refinement
is only allowed in the initial innermost grid that has a comoving side
length of 250 (300) kpc.  We enforce a maximum AMR level of 12 in
these simulation that corresponds to a resolution limit of 2.9 (1.9)
comoving parsecs.

We use the nine species (H, H$^{\rm +}$, He, He$^{\rm +}$, He$^{\rm
  ++}$, e$^{\rm -}$, H$_2$, H$_2^{\rm +}$, H$^{\rm -}$)
non-equilibrium chemistry model in \enzo~\citep{Abel97, Anninos97} and
the \hh~cooling rates from \citet{Galli98}.  Compton cooling and
heating of free electrons by the CMB and radiative losses from atomic
and molecular cooling are computed in the optically thin limit.

We focus on the region containing the most massive halo in the
simulation box.  We perform three calculations -- simulation A with
star formation and radiation transport (RT; SimA-RT), simulation B
with star formation and RT (SimB-RT), and simulation B with star
formation, RT, and SNe (SimB-SN).  We end the calculations at the same
redshift the halo with a virial temperature of 10$^4$ K collapses at z
= 15.9 (16.8) in the hydrogen and helium cooling runs (HHe) of Paper
I.

\subsection{Star Formation Recipe}

Star formation is modelled through an extension \citep{Abel07} of the
\citet{Cen92b} algorithm that automatically forms a star particle when
a grid cell has 
\begin{enumerate}
\item an overdensity exceeding $5 \times 10^5$
\item a converging velocity field ($\nabla \cdot \mathbf{v} < 0$)
\item rapidly cooling gas (\tcool~$<$ \tdyn)
\item an \hh~fraction greater than $5 \times 10^{-4}$.
\end{enumerate}
Then we remove half of the gas from the grid cells in a sphere that
contains twice the stellar mass, which is a free parameter.  Once
these criteria are met, \citet{Abel02} showed that a Pop III star
forms within 10 Myr.  For this reason, we do not impose the Jeans
instability requirement used in \citeauthor{Cen92b} and do not follow
the collapses to stellar scales.  We allow star formation to occur in
the Lagrangian volume of the surrounding region out to three virial
radii from the most massive halo at $z = 10$ in the dark matter only
runs as discussed in Paper I.  This volume that has a side length of
195 (225) comoving kpc at $z = 30$ and 145 (160) comoving kpc at the
end of the calculation.

Runs with only star formation model all Pop III stars with M$_\star$ =
100 \Ms~that live for 2.7 Myr and emit $1.23 \times 10^{50}$ hydrogen
ionizing photons per second.  After its death, the star particle is
converted into an inert 100 \Ms~tracer particle.  The SNe runs use
M$_\star$ = 170 \Ms~that results in a lifetime of 2.3 Myr and $2.57
\times 10^{50}$ ionizing photons per second, in accordance with the no
mass loss stellar models of \citet{Schaerer02}.

When a 170 \Ms~star dies, it injects $E_{\rm{SN}} = 2.7 \times
10^{52}$ erg of thermal energy and 80\feighty~\Ms~of metals, appropriate
for a PISN of a 170 \Ms~star \citep{Heger02}, into a sphere with
radius $r_{\rm{SN}}$ = 1 pc centered on the star's position.  The mass
contained in the star particle and associated metal ejecta are evenly
distributed in this sphere.  The mass of the star particle is changed
to zero, and we track its position in order to determine the number of
stars associated with each halo.  We also evenly deposit the SN energy
in the sphere, which changes the specific energy by
\begin{equation}
  \label{eqn:e_inject}
  \Delta\epsilon = \frac{ \rho_0 \epsilon_0 + \rho_{\rm{SN}}
    \epsilon_{\rm{SN}} } { \rho_0 + \rho_{\rm{SN}} } - \epsilon_0,
\end{equation}
where $\rho_0$ and $\epsilon_0$ are the original gas density and
specific energy, respectively.  Here $\rho_{\rm{SN}} = M_\star /
V_{\rm{SN}}$ is the ejecta density; $\epsilon_{\rm{SN}} = E_{\rm{SN}}
/ M_\star / V_{\rm{SN}}$ is the ejecta specific energy; $V_{\rm{SN}}$
is the volume of a sphere with radius $r_{\rm{SN}}$.  In order not to
create unrealistically strong shocks at the blast wave and for
numerical stability reasons, we smoothly transition from this energy
bubble to the ambient medium, using the function
\begin{equation}
f(r) = A \left\{ 0.5 - 0.5 \tanh \left[ B \left( \frac{r}{r_{\rm{SN}}}
        - 1 \right) \right] \right\}.
\end{equation}
Here $A$ is a normalization factor that ensures $\int f(r) dr = 1$,
and $B$ controls the rate of transition to the ambient medium, where
the transition is steeper with increasing $B$.  We use $A$ = 1.28 and
$B$ = 10 in our calculations.

We continue to use the nine-species chemistry model and do not
consider the additional cooling from metal lines and dust.  We follow
the hydrodynamic transport of metals from the SNe to the enrichment of
the surrounding IGM and halos.

\subsection{Radiative Transfer}

For point sources of radiation, ray tracing is an accurate method to
calculate and evolve radiation fields.  However, millions of rays must
be cast in order to obtain adequate ray sampling at large radii.  We
use adaptive ray tracing \citep{Abel02b} to overcome this dilemma
associated with ray tracing \citep[cf.][]{Abel07}.  We initially cast
768 rays, i.e. level three in HEALPix \citep{Gorski05}, from the
radiation source.  The photons contained in the initial rays are
equal, and their sum is the stellar luminosity.  Rays are split into 4
child rays, whose angles are calculated with the next HEALPix level,
if their associated solid angle is greater than 20\% of the cell area
$(\Delta x)^2$.  Photons are distributed evenly among the children.
This occurs if the ray travels to a large distance from its source, or
the ray encounters a highly refined AMR grid, in which adaptive ray
tracing accurately samples and retains the fine structure contained in
high resolution regions.

The rays cast in these simulations have an energy $E_{\rm{ph}}$ that
is the mean energy of hydrogen ionizing photons from the stellar
source.  For 100 \Ms~and 170 \Ms, this energy is roughly equal at 28.4
and 29.2 eV, respectively, due to the weak dependence of the surface
temperature of primordial stars on stellar mass.  The rays are
transported at the speed of light on timesteps equal to the
hydrodynamical timestep on the current finest AMR level.

The radiation transport is coupled with the hydrodynamical, chemistry,
and energy solvers of \enzo.  Here we only consider hydrogen
photoionization.  We first calculate the photoionization and heating
rates caused by each ray and then sub-cycle the chemistry and heating
solvers with these additional rates on every radiation timestep.  Next
we advance the hydrodynamics of the system with the usual adaptive
timesteps.

The hydrogen photoionization rate is computed by
\begin{equation}
\label{eqn:kph}
k_{ph} = \frac{ P_0 (1 - e^{-\tau}) } { n_{\rm{HI}} \; V_{\rm{cell}} \; \Delta
  t_{{\rm ph}} },
\end{equation}
where $P_0$ is the incoming number of photons, $n_{\rm{HI}}$ is the
number density of neutral hydrogen, $V_{\rm{cell}}$ is the volume of
the computational grid cell, and $\tau = n_{\rm{HI}} \;
\sigma_{\rm{HI}} \; dl$ is the optical depth.  Here $\sigma_{\rm{HI}}$
is the cross section of hydrogen, and $dl$ is the distance travelled by
the ray through the cell.  The heating rate is computed from the
excess photon energies of the photoionizations by
\begin{eqnarray}
\label{eqn:heating}
\Gamma_i &\;=\;& k_{ph} (E_{\rm{ph}} - E_i) \nonumber \\
&=& \frac{ P_0 (1 - e^{-\tau}) } {
  n_{\rm{HI}} \; V_{\rm{cell}} \; \Delta t_{{\rm ph}} } \;
(E_{\rm{ph}} - E_i).
\end{eqnarray}
In the case of hydrogen ionizing photons, $E_i$ = 13.6 eV.  In each
radiation timestep, the number of photons absorbed, i.e. $P_0 (1 -
e^{-\tau})$, is subtracted from the ray.  The ray is eliminated once
most of the associated photons (e.g. 99\%) are absorbed or the ray
encounters a highly optically thick region (e.g. $\tau > 20$).

To model the \hh~dissociating (Lyman-Werner; LW) radiation between
11.2 and 13.6 eV, we use an optically thin 1/$r^2$ radiation field
with luminosities calculated from \citet{Schaerer02}.  We use the
\hh~photo-dissociation rate coefficient for the Solomon process of
$k_{diss} = 1.1 \times 10^8 \flw$ s$^{-1}$, where \flw~is the LW flux
in units of \flux~\citep{Abel97}.  We do not consider the
self-shielding of LW photons.  Because molecular clouds only become
optically thick in the late stages of collapse and above column
densities of $\sim$$10^{14}$ cm$^{-2}$ \citep{Draine96}, we expect our
results to not be drastically affected by neglecting LW
self-shielding.  Additionally, LW self-shielding may be unimportant up
to column densities of $10^{20} - 10^{21}$ cm$^{-2}$ if the medium
contains very large velocity gradients and anisotropies
\citep{Glover01}.

\begin{figure}[t]
\begin{center}
\epsscale{1.15}
\plotone{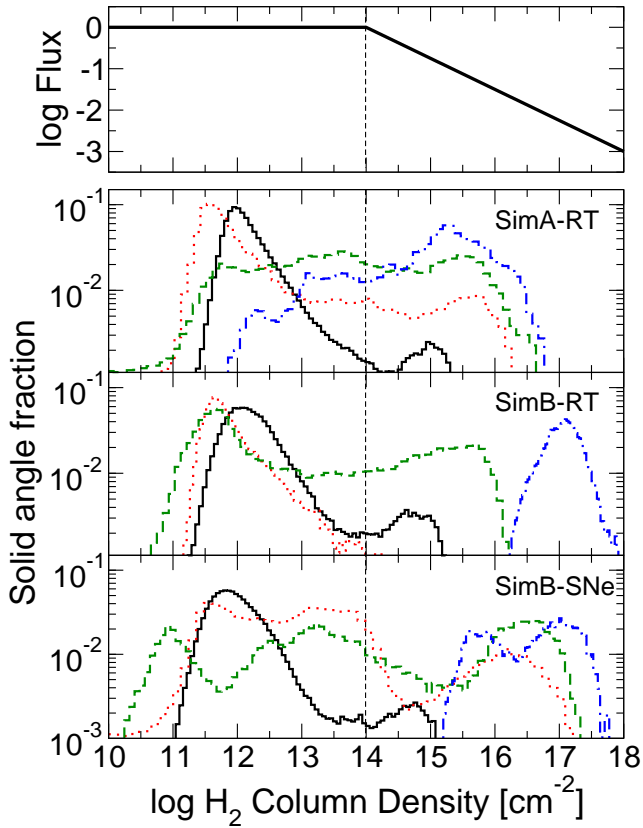}
\caption{\label{fig:lw_tau} \textit{Top:} Normalized, transmitted
  Lyman-Werner flux as a function of \hh~column density
  \citep{Draine96}.  \textit{Bottom:} Histograms of \hh~column density
  at 3 proper kpc from a currently shining star.  The various lines
  are evaulated at different redshifts and host halo masses, which are
  approximately (30, 10$^{5.7} \Ms$) for solid black, (22,
  $10^{6.0}\Ms$) for dotted red, (19, $10^{6.3}\Ms$) for dashed green,
  and (16.8, $10^{7.5}\Ms$) for dash-dotted blue.  The vertical thin
  line at $10^{14}$ cm$^{-2}$ denotes the transition to
  self-shielding.  Halos with $M \gsim 10^6 \Ms$ begin to self-shield
  Lyman-Werner radiation, where most of the LW radiation is
  self-shielded in \tvir~= 10$^4$ K halos.}
\end{center}
\end{figure}

In principle, the propagation of LW photons can be calculated with our
ray tracing software.  However, the computational expense of this
additional coupled calculation is large because the LW radiation field
is essentially optically thin and must be traced through all grid
cells in the simulation.  In comparison, the hydrogen ionizing photons
are only traced within the \ion{H}{2} region whose radii are at most 5
proper kpc.  To test the validity of our approximation, we calculate
the \hh~optical depth from a currently shining star to a spherical
surface with a radius of 3 proper kpc, which is plotted in Figure
\ref{fig:lw_tau}.  The top panel depicts the effect of self-shielding
above a column density of $10^{14}$ cm$^{-2}$ and can be fitted with
the power law, $F = F_0 N_{14}^{-3/4}$.  Here $F$ and $F_0$ are the
transmitted and unobscured flux, respectively, and $N_{14}$ is the
\hh~optical depth in units of $10^{14}$ cm$^{-2}$ \citep{Draine96}.

We calculate the \hh~optical depth in static data outputs in SimA-RT,
SimB-RT, and SimB-SNe with adaptive ray tracing.  The \hh~abundances
are taken from the simulations, which use the optically-thin
assumption.  This causes the optical depths in Figure \ref{fig:lw_tau}
and self-shielding to be underestimated.  Nonetheless, it is useful to
estimate the amount of \hh~self-shielding in these high redshift
halos.  The stars are located in dark matter halos with masses $3
\times 10^7$, $2 \times 10^6$, $10^6$, and $5 \times 10^5$ \Ms~at z
$\approx$ (16.8, 19, 22, 30), respectively.  In halos with $M \lsim
10^6 \Ms$, almost all of the molecular hydrogen within the host and
neighboring halos are not self-shielding, which starts to occur above
column densities of $10^{14}$ cm$^{-2}$.  Above this halo mass, a
fraction of the \hh~column densities rise above this critical value
for \hh~self-shielding.  Thus at lower redshift, we could be
underestimating the star formation rate as overdense clumps may be
shielded to the LW radiation, especially in larger halos or when a
primordial star is living in a neighboring massive halo.

%%%%%%%%%%%%%%%%%%%%%%%%%%%%%%%%%%%%%%%%%%%%%%%%%%%%%%%%%%%%%%%%%%%%%%%%
%
% H2 COLUMN DENSITIES
%
%%%%%%%%%%%%%%%%%%%%%%%%%%%%%%%%%%%%%%%%%%%%%%%%%%%%%%%%%%%%%%%%%%%%%%%%

%%%%%%%%%%%%%%%%%%%%%%%%%%%%%%%%%%%%%%%%%%%%%%%%%%%%%%%%%%%%%%%%%%%%%%%%

\section{RESULTS}
\label{sec:reionResults}

In this section, we first discuss star formation in dwarf galaxy
progenitors.  Then we focus on the global characteristics of the most
massive halo.  We detail the different ISM phases.  Metal transport
from PISNe and the associated metal-enriched star formation history
are discussed last.

\subsection{Number of Star Forming Halos}

Gas in halos with masses $\lsim 10^6$ \Ms~is evacuated by a
$\sim$30\kms~D-type front, leaving a diffuse (1\cubecm) and warm ($3
\times 10^4$ K) medium \citep{Whalen04, Kitayama04, Yoshida06a,
  Abel07}.  The aftermath of SN explosions in relic \ion{H}{2} regions
was explored in spherical symmetry in one-dimensional calculations by
\citet{Kitayama05}.  Even without SNe, star formation is suppressed
for $\sim$100 Myr before gas is reincorporated into the potential well
of this early galactic progenitor.  PISNe provide an extra $\sim
10^{52}$ erg of thermal energy and can evacuate halos up to 10$^7$
\Ms.  Hence star formation within these low-mass halos are highly
dependent on their star formation and merger histories, as illustrated
by \citet{Yoshida06a}.

%%%%%%%%%%%%%%%%%%%%%%%%%%%%%%%%%%%%%%%%%%%%%%%%%%%%%%%%%%%%%%%%%%%%%%%%
%
% SIMULATION SUMMARY
%
%%%%%%%%%%%%%%%%%%%%%%%%%%%%%%%%%%%%%%%%%%%%%%%%%%%%%%%%%%%%%%%%%%%%%%%%

\begin{deluxetable}{lcccc}
%\tablecolumns{5}
\tabletypesize{}
\tablewidth{0pc}
\tablecaption{Global halo properties with star formation\label{tab:sf}}

\tablehead{ \colhead{Name} & \colhead{N$_\star$($<$\rvir)} & 
  \colhead{N$_\star$($<$3\rvir)} & 
  \colhead{M$_{\rm{gas}}$/M$_{\rm{tot}}$} & \colhead{$\lambda_{\rm{g}}$}
}
\startdata
SimA-HHe & \dots & \dots & 0.14  & 0.010 \\
SimA-RT  & 11    & 12    & 0.064 & 0.030 \\
SimB-HHe & \dots & \dots & 0.14  & 0.010 \\
SimB-RT  & 22    & 26    & 0.089 & 0.014 \\
SimB-SN  & 4     & 10    & 0.046 & 0.038 
\enddata

\tablecomments{Col. (1): Simulation name. Col. (2): Number of stars
  hosted in the halo and its progenitors. Col. (3): Number of stars
  formed in the Lagrangian volume within 3\rvir~of the most massive
  halo. Col. (4): Baryon fraction within \rvir. Col. (5): Baryonic
  spin parameter [Eq. \ref{eqn:spin}].}

\end{deluxetable}
%%%%%%%%%%%%%%%%%%%%%%%%%%%%%%%%%%%%%%%%%%%%%%%%%%%%%%%%%%%%%%%%%%%%%%%%

Table \ref{tab:sf} summarizes the global properties of the most
massive halo the time of collapse in the HHe calculations.
Approximately 5--20 Pop III stars form in the progenitors, whose
original gas structures are nearly destroyed by radiative feedback, of
the 10$^4$ K halo.  More specifically at z = 15.9 (16.8), there are
19, 29, and 24 stars that form after redshift 30 in the SimA-RT,
SimB-RT, and SimB-SN runs, respectively.  The most massive halo and
its progenitors have hosted 11, 22, and 4 stars in the same
simulations.  When we look at the Lagrangian volume contained within
three times the virial radius of the most massive halo, there have
been 12, 26, and 10 instances of star formation.  For comparison,
\citet{Greif08} find an an upper limit of 10 Population III stars that
are hosted in the progenitors of an early galaxy similar to the ones
presented in this work.

\subsection{Global Nature of Objects}

\begin{figure}[t]
\begin{center}
\epsscale{1.15}
\plotone{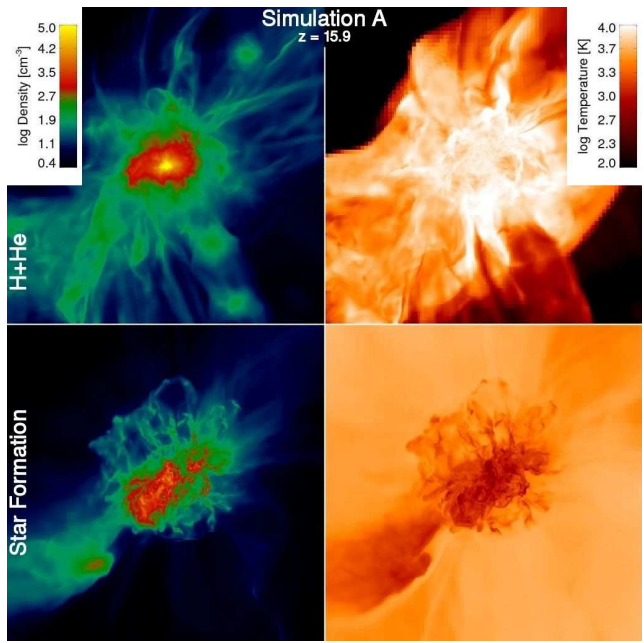}
\caption{\label{fig:proj_a} Density-squared weighted projections of
  gas density (\textit{left}) and temperature (\textit{right}) of the
  most massive halo in simulation A.  The field of view is 1.2 proper
  kpc.  The \textit{top} row shows the model without star formation
  and only atomic hydrogen and helium cooling.  The \textit{bottom}
  row shows the same halo affected by primordial star formation.  Note
  the filamentary density structures, clumpy interstellar medium, and
  the counter-intuitive effect that feedback leads to lower
  temperatures.}
\end{center}
\end{figure}

\begin{figure}[t]
\begin{center}
\epsscale{1.15}
\plotone{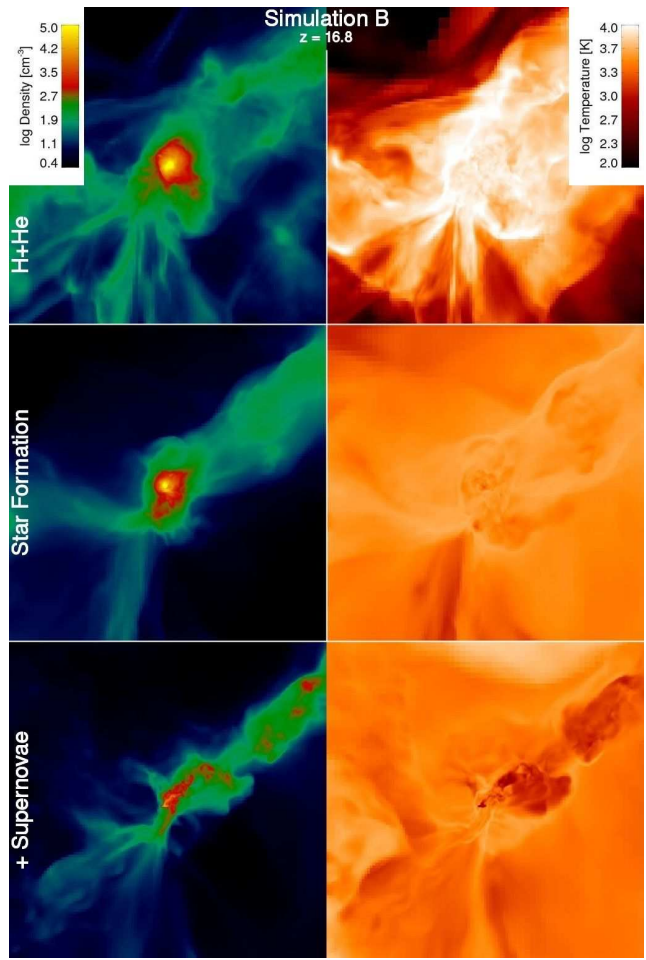}
\caption{\label{fig:proj_b} Same as Figure \ref{fig:proj_a} for
  simulation B.  Here the \textit{bottom} row shows the halo with
  primordial stellar feedback and supernovae.}
\end{center}
\end{figure}

In addition to providing the first ionizing photons and metals to the
universe, Pop III stars change the global gas dynamics of \tvir~$\sim$
10$^4$ K star forming halos.  Figures \ref{fig:proj_a} and
\ref{fig:proj_b} compare the structure of the most massive halo in all
simulations, depicting density-squared weighted projections of gas
density and temperature.  All of the halos have a virial mass of $3.5
\times 10^7 \Ms$.  The models with neither star formation nor
\hh~chemistry show a centrally concentrated, condensing \tvir~= 10$^4$
K halo with its associated virial heating.  In comparison, feedback
from primordial star formation expels the majority of the gas in
low-mass star forming progenitors.  The combination of these outflows
and accretion from filaments induces the formation of an inhomogeneous
medium, where the radiation anisotropically propagates, creating
champagne flows in the directions with lower column densities.  The
temperature projections illustrate both the ultraviolet heated
($\sim10^4$ K) and optically thick, cool ($\sim 10^3$ K) regions in
the host halo and IGM.  Also note the nearby substructure is
photo-evaporated by the stars hosted in the most massive progenitor.
SN explosions alter the gas structure by further stirring and ejecting
material after main sequence.  Next we quantify these visual features
with the baryon fraction of the halo and an inspection of phase
diagrams of density and temperature.

%%%%%%%%%%%%%%%%%%%%%%%%%%%%%%%%%%%%%%%%%%%%%%%%%%%%%%%%%%%%%%%%%%%%%%%%
%
% SIM A PROJECTIONS
%
%%%%%%%%%%%%%%%%%%%%%%%%%%%%%%%%%%%%%%%%%%%%%%%%%%%%%%%%%%%%%%%%%%%%%%%%

%%%%%%%%%%%%%%%%%%%%%%%%%%%%%%%%%%%%%%%%%%%%%%%%%%%%%%%%%%%%%%%%%%%%%%%%

Because the gas in the progenitors is mostly evacuated, the baryon
fraction of high-redshift star forming halos are greatly reduced.  In
halos with masses $\lsim 10^6$ \Ms, the baryon fraction lowers to $5
\times 10^{-3}$ ten million years after the star's death without a SN
\citep[cf.][]{Yoshida06a}.  When we include SNe, the baryon fraction
decreases further to $1 \times 10^{-5}$ six million years after the
explosion \citep[cf.][]{Kitayama05}.  This is in stark contrast with
the cosmic fraction \Ob/\Om~= 0.17.

Within these shallow potential wells, outflows from stellar feedback
impede subsequent star formation until sufficient gas is
reincorporated, occurring through mergers and smooth IGM accretion.
After the halo mass surpasses $\sim$$3 \times 10^6 \Ms$, total
evacuation does not occur but significant outflows are still
generated.  Near the same mass scale, multiple sites of star formation
occur in the same halo.  In our simulations, these stars rarely shine
simultaneously since their lifetimes is significantly shorter than the
Hubble timescale on which the halos are assembled.  We neglect
\hh~self-shielding, which may affect the timing of star formation but
unlikely not the global star formation rate \citep{Ahn07}.  Here the
nearby overdensity survives the blast of UV photo-heating from the
previous star and condenses to form a star a few Myr afterwards.  This
scenario of adjacent star formation is similar to the one presented in
\citet{Abel07}, but the multiple sites of star formation are caused by
\hh~and \lya~cooling in central protogalactic shocks \citep{Shapiro87}
and aided by additional \hh~cooling in ionization fronts
\citep{Ricotti01, Ahn07, Whalen08A}, not from residual cores from a
recent major merger.

%%%%%%%%%%%%%%%%%%%%%%%%%%%%%%%%%%%%%%%%%%%%%%%%%%%%%%%%%%%%%%%%%%%%%%%%
%
% SIM B PROJECTIONS
%
%%%%%%%%%%%%%%%%%%%%%%%%%%%%%%%%%%%%%%%%%%%%%%%%%%%%%%%%%%%%%%%%%%%%%%%%

%%%%%%%%%%%%%%%%%%%%%%%%%%%%%%%%%%%%%%%%%%%%%%%%%%%%%%%%%%%%%%%%%%%%%%%%

%%%%%%%%%%%%%%%%%%%%%%%%%%%%%%%%%%%%%%%%%%%%%%%%%%%%%%%%%%%%%%%%%%%%%%%%
%
% BARYON FRACTION AND SPIN PARAMETER
%
%%%%%%%%%%%%%%%%%%%%%%%%%%%%%%%%%%%%%%%%%%%%%%%%%%%%%%%%%%%%%%%%%%%%%%%%
\begin{figure}[b]
\begin{center}
\epsscale{1.15}
\plotone{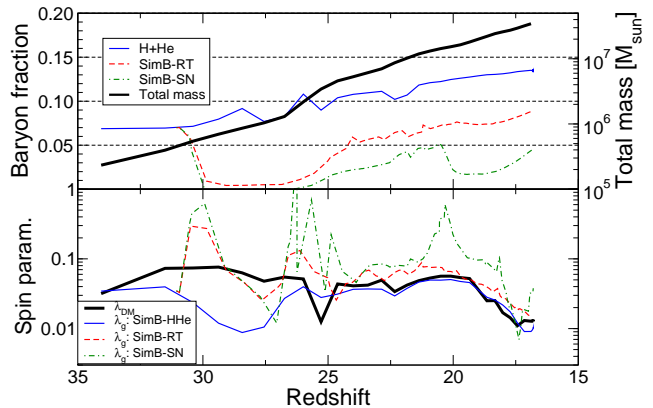}
\caption{\label{fig:spinmass} The top graph shows the total mass
  (thick black line) of the most massive progenitor in simulation B.
  The thin lines depict the baryon fraction, $M_{\rm{gas}} /
  M_{\rm{tot}}$, of simulation B without star formation (blue solid), with
  star formation (red dashed), and with SNe (green dot-dashed).  The
  bottom graph shows the spin parameter (eq. \ref{eqn:spin}) of the
  same halo in the dark matter (thick line) and gas (thin lines, same
  legend as top graph).}
\end{center}
\end{figure}
%%%%%%%%%%%%%%%%%%%%%%%%%%%%%%%%%%%%%%%%%%%%%%%%%%%%%%%%%%%%%%%%%%%%%%%%

After the first star forms in the most massive progenitor, the halo
slowly regains gas mass mainly through merger events of halos that
have not experienced star formation.  We show the evolution of the
baryon fraction of simulation B in the top section of Figure
\ref{fig:spinmass}.  In SimB-RT, ionization fronts created from stars
in halos with $\mvir \gsim 3 \times 10^6 \Ms$ cannot expel the
majority of baryons in the halo but only generate outflows.  In
SimB-SN, similar recovery occurs but the additional energy from the
three SNe at redshift 20 evacuates the halo once again to a baryon
fraction of 0.02.  When the most massive halo reaches \tvir~$\sim
10^4$ K, the baryon fraction within the virial radius has only
partially recovered to 0.064, 0.089, and 0.046 in the SimA-RT,
SimB-RT, and SimB-SN calculations.  Without any stellar feedback,
these fractions are 0.14 in the HHe runs.

%%%%%%%%%%%%%%%%%%%%%%%%%%%%%%%%%%%%%%%%%%%%%%%%%%%%%%%%%%%%%%%%%%%%%%%%
%
% DARK MATTER PROFILES
%
%%%%%%%%%%%%%%%%%%%%%%%%%%%%%%%%%%%%%%%%%%%%%%%%%%%%%%%%%%%%%%%%%%%%%%%%
\begin{figure}[t]
\begin{center}
\epsscale{1.15}
\plotone{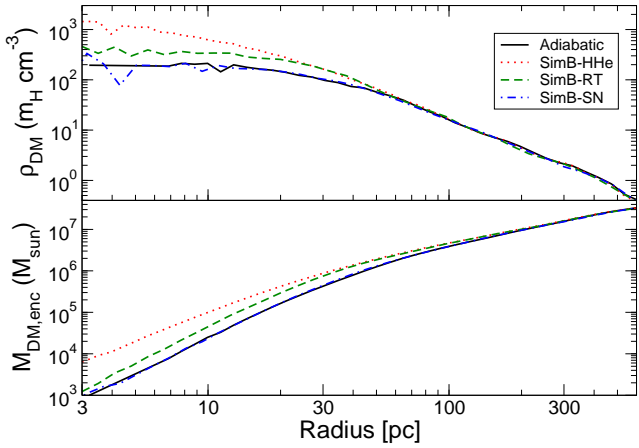}
\caption{\label{fig:dm} Radial profiles of DM density (top panel) and
  enclosed DM mass (bottom panel) of the most massive halo in
  simulation B at $z$ = 16.8 with an adiabatic equation of state
  (solid), no star formation (dotted), star formation and feedback
  (dashed), and plus SN feedback (dash-dotted).  The reduced central
  baryon overdensities in the progenitors lead to lower central DM
  densities.}
\end{center}
\end{figure}
%%%%%%%%%%%%%%%%%%%%%%%%%%%%%%%%%%%%%%%%%%%%%%%%%%%%%%%%%%%%%%%%%%%%%%%%

During central gaseous collapses in halos, the gravitational potential
in the inner $\sim$50 pc becomes baryon dominated where the baryon
density is greater than the DM density, leading to a contraction of
the DM inner halo \citep{Blumenthal86, Gnedin04}.  If stellar feedback
evaporates this overdensity, the central potential will not be as deep
during the assembly of early dwarf galaxies, possibly resulting in the
DM being not as centrally concentrated.  We plot radial profiles of DM
density and enclosed DM mass of the most massive halo in simulation B
at $z$ = 16.8 in Figure~\ref{fig:dm}.  The dark matter density is
decreased in the inner 20 pc (0.03\rvir) up to a factor of 5 with
stellar feedback alone.  The effect is exacerbated by the additional
feedback from SN explosions inside 50 pc (0.1\rvir), and the central
DM densities decrease by another factor of 2, similar to adiabatic
calculations.

These outflows also create inhomogeneities in and around halos and
increase the baryonic spin parameter
\begin{equation}
\label{eqn:spin}
\lambda_{\rm{g}} = \frac{L_{\rm{g}} \vert E_{\rm{g}}
  \vert^{1/2}}{GM_{\rm{g}}^{5/2}},
\end{equation}
where $L_{\rm{g}}$, $E_{\rm{g}}$, and $M_{\rm{g}}$ are the total
baryonic angular momentum, kinetic energy, and mass of the system
\citep{Peebles71}.  This is basically the ratio of rotational to
gravitational energy of the system.  The DM spin parameter $\lambda$
uses the total DM angular momentum, kinetic energy, and mass of the
system.

Interestingly the halo experiences significant fluctuations in
$\lambda_g$ when significant outflows are generated by stellar
feedback.  We plot its evolution in the most massive progenitor in the
bottom section of Figure \ref{fig:spinmass}.  Without feedback (HHe)
after redshift 25, the gas and DM spin parameters are approximately
equal and follow the same trends.  However when star formation
(SimB-RT) is included, the spin parameter is increased by an order of
magnitude after the first star.  It then decays over the next 40 Myr
but increases again after the second star in the halo at redshift 26.
$\lambda_g$ continues to be up to a factor of two higher than without
star formation after the star can no longer expel most of the gas from
the halo.  In SimB-SN, these effects are even more apparent,
especially at redshift 20 during an episode of three SNe evacuating
the most of the gas from the halo.

At the time of collapse in the HHe runs, $\lambda$ = 0.022 (0.013) and
is marginally lower than the average $\langle \lambda \rangle \simeq
0.04$ found in numerical simulations \citep{Barnes87, Eisenstein95}.
It is not affected by stellar feedback as DM dominates the potential
well.  Without star formation, the baryonic spin parameter
$\lambda_{\rm{g}}$ = 0.010 (0.010) and is slightly lower than
$\lambda$.  However with stellar and SNe feedback, $\lambda_{\rm{g}}$
increases up to a factor of 4.  The effect is smaller without SNe but
still significant, raising $\lambda_{\rm{g}}$ to 0.030 (0.014).

The increase of $\lambda_{\rm{g}}$ could be caused by the forces
generated in the \ion{H}{2} region and SNe blastwaves \citep{Abel01}.
Because these events can expel gas from the potential well and
$\langle \lambda \rangle \simeq 0.04$, these forces can be up to 25
times greater than the cosmological tidal torques usually associated
with the angular momentum of galaxies \citep{Peebles69}.  The angular
momentum associated with ionization fronts and blastwaves would have
to be almost opposite of angular momentum vector of the halo to slow
the rotation.  Therefore we expect these outflows to produce on
average an overall increase the angular momentum of galaxies.  In the
assembly of early dwarf galaxies, a fraction of the expelled gas
experiencing large scale torques longer falls back, now with a higher
specific angular momentum, to the galaxy and increasing its spin
parameter further.

%%%%%%%%%%%%%%%%%%%%%%%%%%%%%%%%%%%%%%%%%%%%%%%%%%%%%%%%%%%%%%%%%%%%%%%%
%
% RADIAL PROFILES
%
%%%%%%%%%%%%%%%%%%%%%%%%%%%%%%%%%%%%%%%%%%%%%%%%%%%%%%%%%%%%%%%%%%%%%%%%

\begin{figure}[t]
\begin{center}
\epsscale{1.15}
\plotone{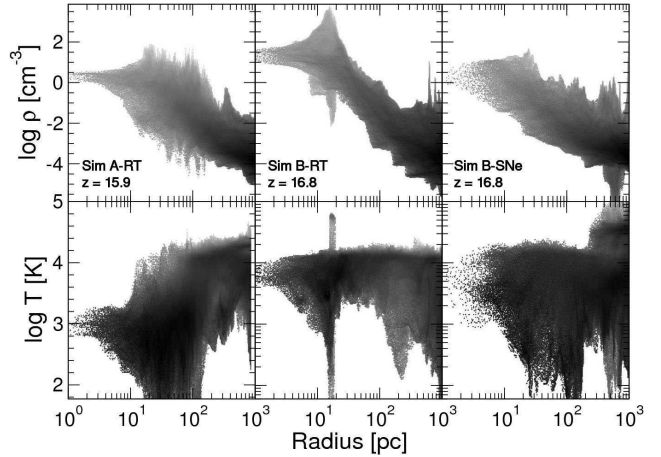}
\caption{\label{fig:radial} Mass-weighted radial profiles of density
  (\textit{top}) and temperature (\textit{bottom}), centered on the
  densest DM particle.  The columns show data from SimA-RT
  (\textit{left}), SimB-RT (\textit{middle}), and SimB-SN
  (\textit{right}).  Note how the cool and warm gas phases coexist at
  similar radii throughout the halo.}
\end{center}
\end{figure}
%%%%%%%%%%%%%%%%%%%%%%%%%%%%%%%%%%%%%%%%%%%%%%%%%%%%%%%%%%%%%%%%%%%%%%%%

%%%%%%%%%%%%%%%%%%%%%%%%%%%%%%%%%%%%%%%%%%%%%%%%%%%%%%%%%%%%%%%%%%%%%%%%
%
% PHASE DIAGRAMS
%
%%%%%%%%%%%%%%%%%%%%%%%%%%%%%%%%%%%%%%%%%%%%%%%%%%%%%%%%%%%%%%%%%%%%%%%%
\begin{figure}[t]
\begin{center}
\epsscale{1.15}
\plotone{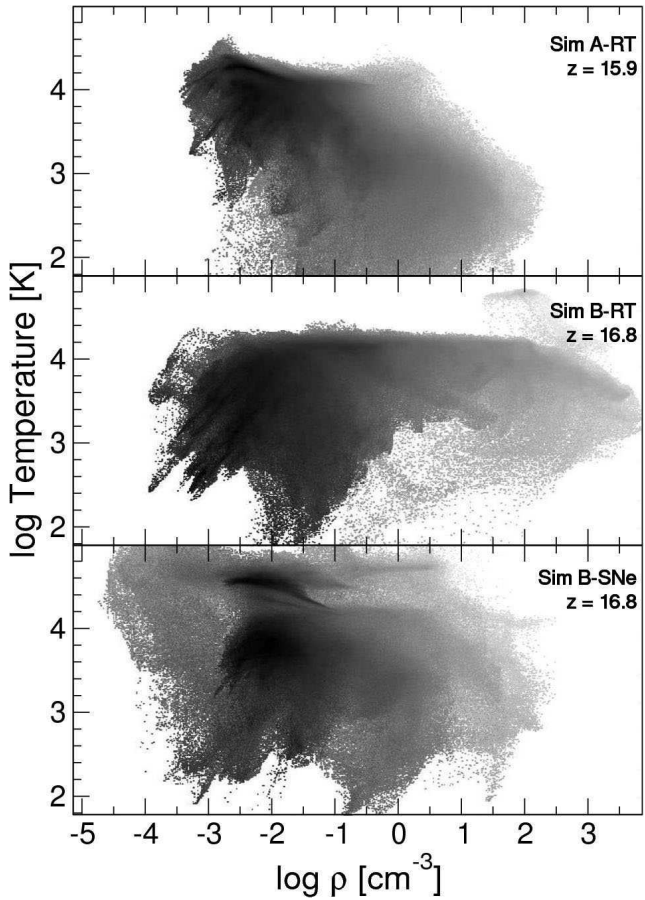}
\caption{\label{fig:phase} Mass-weighted $\rho$--T phase diagrams of a
  sphere with radius 1 kpc, centered on the most massive halo in
  SimA-RT (\textit{top}), SimB-RT (\textit{middle}), and SimB-SN
  (\textit{bottom}).  At $T > 10^4$ K, one can see the \ion{H}{2}
  regions created by current star formation.  The warm, low density
  ($\rho < 10^{-3} \cubecm$) gas in SimB-SN are contained in SNe
  shells.}
\end{center}
\end{figure}
%%%%%%%%%%%%%%%%%%%%%%%%%%%%%%%%%%%%%%%%%%%%%%%%%%%%%%%%%%%%%%%%%%%%%%%%

\subsection{ISM Phases}

The combination of molecular cooling, stellar feedback, and SN
explosions create a multi-phase ISM in star forming halos.  These
phases are interspersed throughout the halo.  They are marginally seen
in the temperature projections in Figures~\ref{fig:proj_a} and
\ref{fig:proj_b}.  However they are better demonstrated by the
mass-weighted radial profiles in Figure~\ref{fig:radial} and
density-temperature phase diagrams in Figure~\ref{fig:phase}.  The
radial profiles are centered on the densest DM particle. For a given
radius within the halo, the gas density can span up to 6 orders of
magnitude, and the temperature ranges from $\sim$100 K in the cool
phase to 30,000 K in the warm, ionized phase.  Below we describe the
different ISM phases at redshift 15.9 and 16.8 for simulation A and B,
respectively.

\medskip

\textit{Cool phase}--- The relatively dense ($\rho > 100 \cubecm$) and
cool ($T > 1000$ K) gas has started to condense by \hh~cooling.
Current star formation dissociates \hh~in nearby condensations through
LW radiation in our simulations.  In many cases, especially when \mvir
$\gsim 10^7$ \Ms, nearby clumps remain cool and optically thick.
After the star dies, \hh~formation can proceed again to form a star in
these clumps.  There are two other sources of cool gas.  First, the
filaments are largely shielded from being photo-heated and provide the
galaxy with cold accretion flows.  Second, after the main sequence,
the material within the expanding shell either from a D-type front or
SN blastwave cools through adiabatic expansion and Compton cooling to
temperatures as low as 100 K, which is seen in the $\rho$-T phase
diagram at very low densities.

\textit{Warm, neutral phase}--- Gas that cools by atomic hydrogen line
transitions, but not molecular, has $T \sim 8000$ K and densities
ranging from $10^{-3}$ to 10$^2 \cubecm$.  Gas in relic \ion{H}{2}
regions and virially shock-heated gas compose this phase.

\textit{Warm, ionized phase}--- In the SimB-RT simulation at the final
redshift, a Pop III star is creating an \ion{H}{2} region with
temperatures up to 30,000~K.  It is 20 pc from the halo center.  The
\ion{H}{2} region does not breakout from this \tvir~= 10$^4$ K halo.

\begin{figure}[t]
\begin{center}
\epsscale{1.15}
\plotone{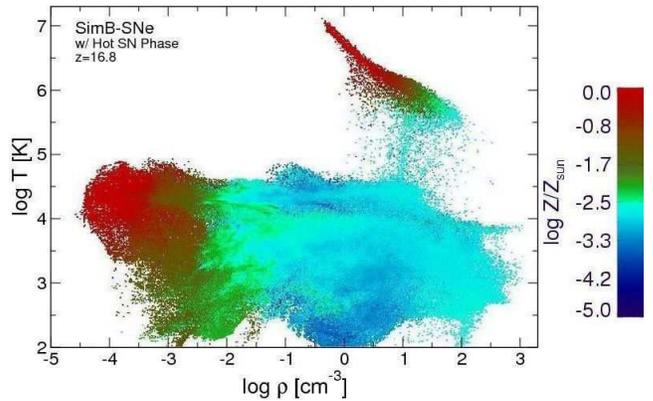}
\caption{\label{fig:hotphase} The same phase diagram of SimB-SN in
  Figure \ref{fig:phase} but colored by mean metallicity and scaled to
  show the hot, X-ray phase.  These data are 45 kyr after a
  pair-instability SN.  The SN remnants that are warm and diffuse
  ($\rho < 10^{-3} \cubecm$) have solar metallicities or greater.  The
  majority of the ISM has metallicities $\sim$10$^{-2.5}$ solar.  The
  densest, collapsing material has a metallicity $\sim$10$^{-3.5}$
  solar.}
\end{center}
\end{figure}

\textit{Hot, X-ray phase}--- The $3 \times 10^{52}$ ergs of energy
deposited by one PISN in the SimB-SN simulation heats the gas to over
$10^8$ K immediately after the explosion.  Figure \ref{fig:hotphase}
shows the density and temperature of the ISM 45 kyr after a SN, where
the adiabat of the hot phase is clearly visible at $T > 10^5$ K.  A
blastwave initially travelling at 4000\kms~sweeps through the ambient
medium during the free expansion phase.  The gas behind the shock
cools adiabatically and through Compton cooling as the SN shell
expands.

\medskip

%%%%%%%%%%%%%%%%%%%%%%%%%%%%%%%%%%%%%%%%%%%%%%%%%%%%%%%%%%%%%%%%%%%%%%%%
%
% HOT SN PHASE DIAGRAM WITH METALLICITY
%
%%%%%%%%%%%%%%%%%%%%%%%%%%%%%%%%%%%%%%%%%%%%%%%%%%%%%%%%%%%%%%%%%%%%%%%%

%%%%%%%%%%%%%%%%%%%%%%%%%%%%%%%%%%%%%%%%%%%%%%%%%%%%%%%%%%%%%%%%%%%%%%%%

%%%%%%%%%%%%%%%%%%%%%%%%%%%%%%%%%%%%%%%%%%%%%%%%%%%%%%%%%%%%%%%%%%%%%%%%
%
% MACH NUMBERS
%
%%%%%%%%%%%%%%%%%%%%%%%%%%%%%%%%%%%%%%%%%%%%%%%%%%%%%%%%%%%%%%%%%%%%%%%%
\begin{figure}[b]
\begin{center}
\epsscale{1.15}
\plotone{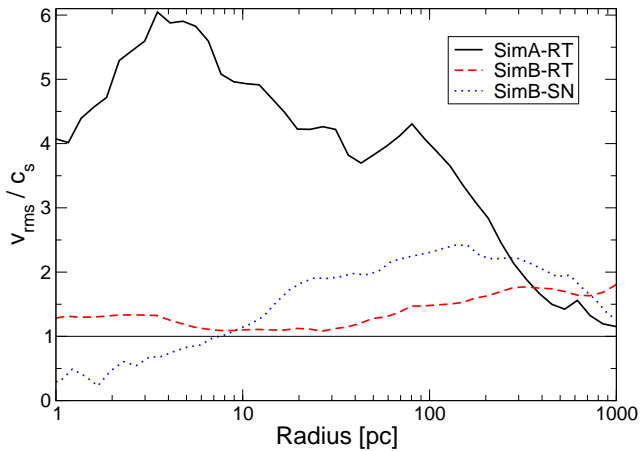}
\caption{\label{fig:mach} Turbulent Mach numbers for SimA-RT (solid),
  SimB-RT (dashed), and SimB-SN (dotted).  The Mach numbers in
  simulation B are lower than simulation A because of the higher
  temperatures created by stellar feedback that occurred shortly
  before the displayed data.}
\end{center}
\end{figure}
%%%%%%%%%%%%%%%%%%%%%%%%%%%%%%%%%%%%%%%%%%%%%%%%%%%%%%%%%%%%%%%%%%%%%%%%

\citet{Wise07a} investigated the generation of turbulence during
virialization without stellar feedback.  The supersonic nature of
turbulence in these halos remains even with Pop III stellar feedback
and \hh~cooling.  In Figure~\ref{fig:mach}, we plot the turbulent Mach
number, $v_{rms} / c_s$, in the most massive halo.  Here $v_{rms}$ is
the three-dimensional rms velocity relative to the mean velocity of
each spherical shell, and $c_s$ is the sound speed.  Turbulent Mach
numbers reach 6 in SimA-RT but is only 1--2 in SimB-RT and SimB-SN,
where the recent stellar feedback in the halo has photo-heated the
gas.  In contrast, sufficient time has elapsed since the previous
episode of star formation in SimA-RT, thus allowing the gas to cool by
\hh~to $\sim$300 K.  Clearly these values are dependent on the stellar
feedback within the halo and possibly its merger history, hence this
range may be representative of turbulent Mach numbers one would find
in early dwarf galaxies.

\subsection{Metallicity}

\begin{figure}[t]
\begin{center}
\epsscale{1.15}
\plotone{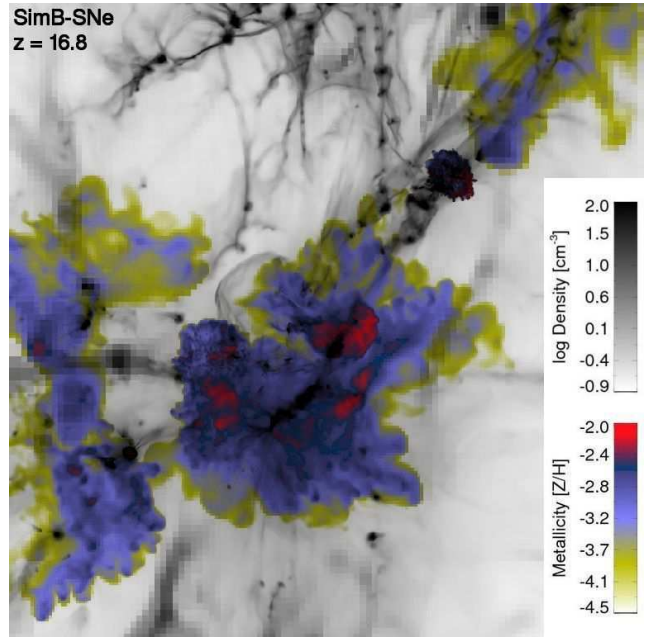}
\caption{\label{fig:metals_154kpc} Density-squared weighted
  projections of metallicity (\textit{color}) and gas density
  (\textit{black and white}) for SimB-SNe at $z = 16.8$ of the inner
  8.6 proper kpc of the simulation.}
\end{center}
\end{figure}

\begin{figure}[t]
\begin{center}
\epsscale{1.15}
\plotone{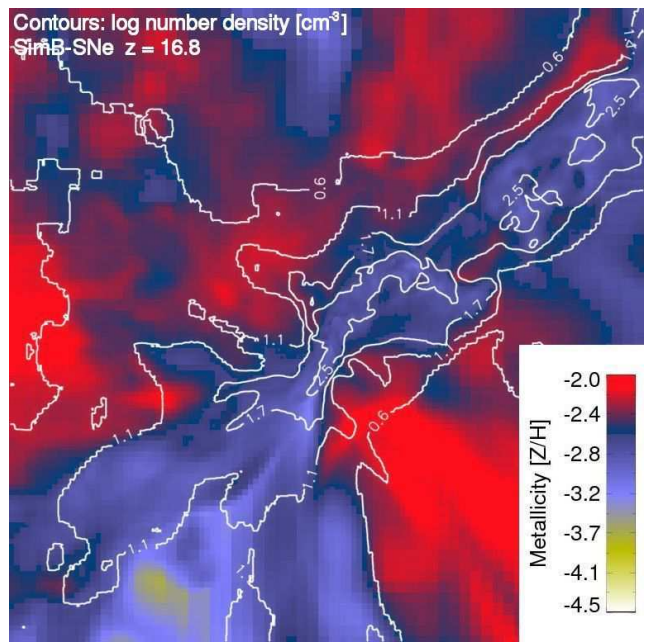}
\caption{\label{fig:metals_rvir} Density-squared weighted projections
  of metallicity, centered on the most massive halo in SimB-SNe at $z
  = 16.8$.  The contours depict the number density of baryons for $n$
  = (4, 13, 50, 320) \cubecm.  The field of view is 1.2 proper kpc and the
  same as Figure \ref{fig:proj_b}.  The projection shows data in a
  slab that are 1.2 proper kpc thick.}
\end{center}
\end{figure}

Metallicities of second and later generations of stars depend on the
location of previous SNe.  Figure \ref{fig:metals_154kpc} shows a
projection of metallicity for the inner 8.7 proper kpc of the SimB-SNe
at $z = 16.8$, which is overplotted against the density-squared
weighted projection of gas density.  We also show the metallicity
projection for the region surrounding the most massive halo in Figure
\ref{fig:metals_rvir}.  This projection depicts the data in a cube
with a side of 1.2 proper kpc, centered on the halo.  The contours in
this figure mark the number density of gas.  In the SimB-SN
calculation, outflows carry most of the SN ejecta to radii up to
$\sim$1 proper kpc after 30 Myr.  Interestingly they approximately
fill the relic \ion{H}{2} region and expand little beyond that.  The
low density IGM marginally resists the outflows, and it is
preferentially enriched instead of the surrounding filaments and
halos.

%%%%%%%%%%%%%%%%%%%%%%%%%%%%%%%%%%%%%%%%%%%%%%%%%%%%%%%%%%%%%%%%%%%%%%%%
%
% METALLICITY PROJECTION (LARGE-SCALE)
%
%%%%%%%%%%%%%%%%%%%%%%%%%%%%%%%%%%%%%%%%%%%%%%%%%%%%%%%%%%%%%%%%%%%%%%%%
%%%%%%%%%%%%%%%%%%%%%%%%%%%%%%%%%%%%%%%%%%%%%%%%%%%%%%%%%%%%%%%%%%%%%%%%

%%%%%%%%%%%%%%%%%%%%%%%%%%%%%%%%%%%%%%%%%%%%%%%%%%%%%%%%%%%%%%%%%%%%%%%%
%
% METALLICITY PROJECTION (SMALL-SCALE)
%
%%%%%%%%%%%%%%%%%%%%%%%%%%%%%%%%%%%%%%%%%%%%%%%%%%%%%%%%%%%%%%%%%%%%%%%%
%%%%%%%%%%%%%%%%%%%%%%%%%%%%%%%%%%%%%%%%%%%%%%%%%%%%%%%%%%%%%%%%%%%%%%%%

It should be noted that this calculation is an upper limit of
metallicities since all stars end with a PISN.  The mixing and
transport of the first metals is a fundamental element of the
transition to Pop II stars and is beneficial to study in detail.  All
metallicities quoted are in units of solar metallicity.  The
metallicities also scale approximately linearly with metal yield of
each SN because we treat the metal field as a tracer field that is
advected with the fluid flow.  We quote the metallicities according to
this scaling.

\begin{figure}[b]
\begin{center}
\epsscale{1.15}
\plotone{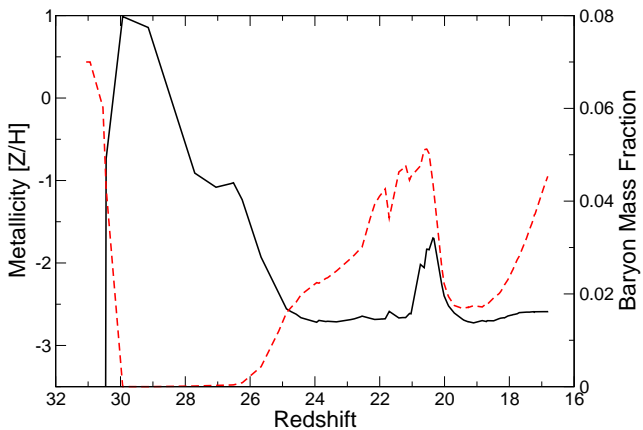}
\caption{\label{fig:metals_evo} Mean metallicity (solid) and baryon
  fraction (dashed) in the most massive progenitor in SimB-SN.  The
  first star enriches the diffuse gas above solar metallicity, which
  then decreases as pristine gas falls into the halo through accretion
  and mergers.  Notice that the metallicity stays roughly constant at
  \tento{-3} when no star formation is occurring.}
\end{center}
\end{figure}

We plot the mean metallicity of the most massive progentior of this
dwarf galaxy in Figure \ref{fig:metals_evo} as a function of redshift.
The first star enriches the gas depleted host halo, and the
metallicity decreases as metal-free gas is incorporated through
mergers and IGM accretion.  The temporary increase in metallicity at
$z = 20$ is associated with the three stars that form in succession.

%%%%%%%%%%%%%%%%%%%%%%%%%%%%%%%%%%%%%%%%%%%%%%%%%%%%%%%%%%%%%%%%%%%%%%%%
%
% METALLICITY EVOLUTION
%
%%%%%%%%%%%%%%%%%%%%%%%%%%%%%%%%%%%%%%%%%%%%%%%%%%%%%%%%%%%%%%%%%%%%%%%%
%%%%%%%%%%%%%%%%%%%%%%%%%%%%%%%%%%%%%%%%%%%%%%%%%%%%%%%%%%%%%%%%%%%%%%%%

%%%%%%%%%%%%%%%%%%%%%%%%%%%%%%%%%%%%%%%%%%%%%%%%%%%%%%%%%%%%%%%%%%%%%%%%
%
% METAL COOLING RATES
%
%%%%%%%%%%%%%%%%%%%%%%%%%%%%%%%%%%%%%%%%%%%%%%%%%%%%%%%%%%%%%%%%%%%%%%%%
\begin{figure*}[t]
\begin{center}
\epsscale{1.15}
\plotone{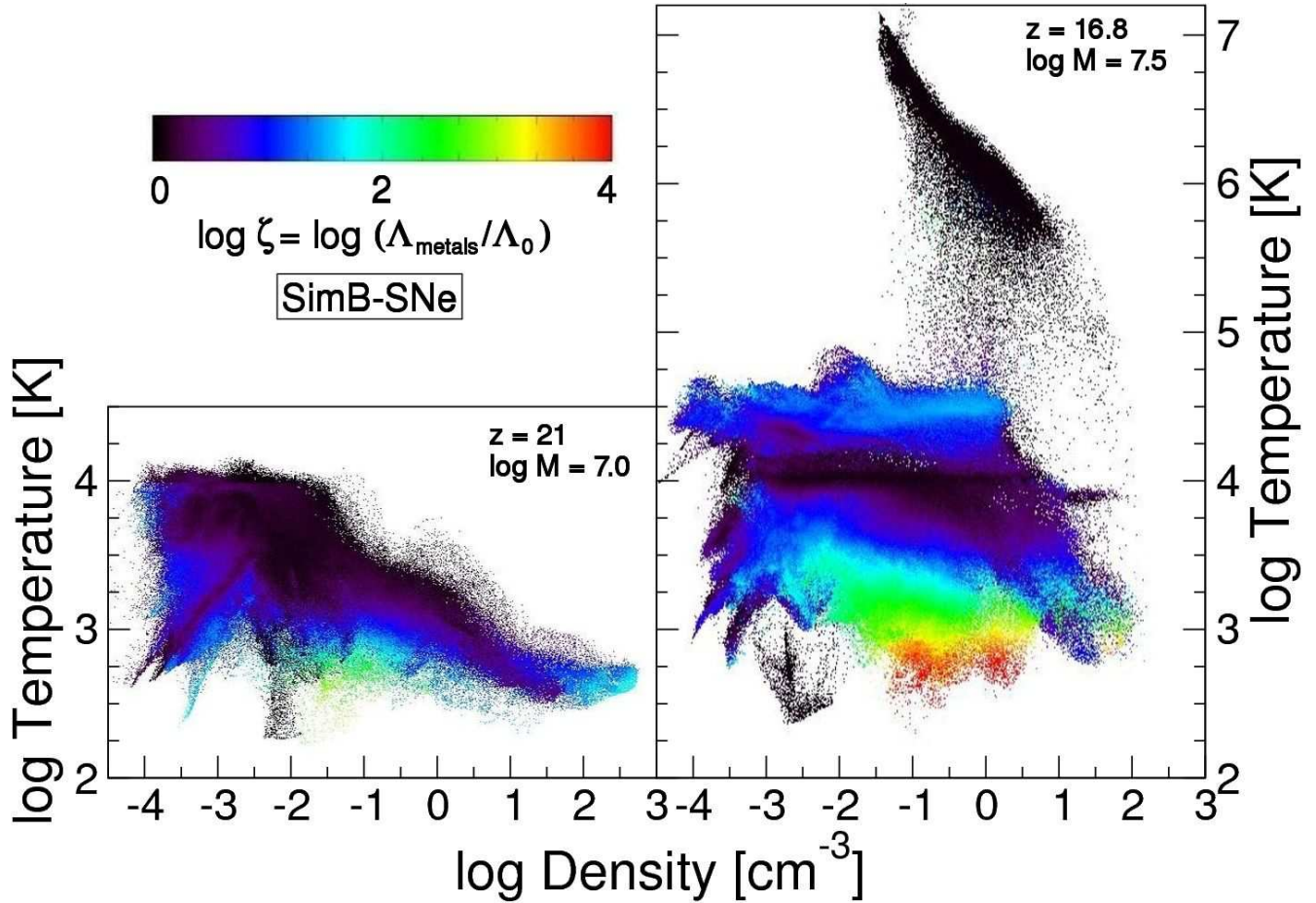}
\caption{\label{fig:metalcool} Mass-weighted $\rho$--T phase diagrams
  of a sphere with radius 1 kpc, centered on the most massive halo in
  SimB-SN at redshift 21 (left) and 16.8 (right).  The halo mass at z
  = (21, 16.8) is $1.1 \times 10^7 \Ms$ and $2.9 \times 10^7 \Ms$,
  respectively.  The pixel intensity is determined by the average of
  the ratio $\zeta$ of the cooling rate with metal-line cooling and
  the cooling rate with primordial chemistry.  At $z = 21$, we
  underestimate cooling rates in gas with $T \gsim 1000$ K and $\rho
  \sim 0.1 \cubecm$ up to a factor of 10, which only composes
  $\sim$2\% of the halo gas mass.  However, as the gas becomes more
  enriched, metal-line cooling becomes dominant over H, He, and
  \hh~cooling in cool gas and old SN remnants ($T > 10^4$ K, $\rho <
  10^{-3} \cubecm$).  However, the latter has a cooling time greater
  than the Hubble time.}
\end{center}
\end{figure*}
%%%%%%%%%%%%%%%%%%%%%%%%%%%%%%%%%%%%%%%%%%%%%%%%%%%%%%%%%%%%%%%%%%%%%%%%

%%%%%%%%%%%%%%%%%%%%%%%%%%%%%%%%%%%%%%%%%%%%%%%%%%%%%%%%%%%%%%%%%%%%%%%%
%
% METAL COOLING RATES
%
%%%%%%%%%%%%%%%%%%%%%%%%%%%%%%%%%%%%%%%%%%%%%%%%%%%%%%%%%%%%%%%%%%%%%%%%
\begin{figure*}
\begin{center}
\epsscale{1.15}
\plottwo{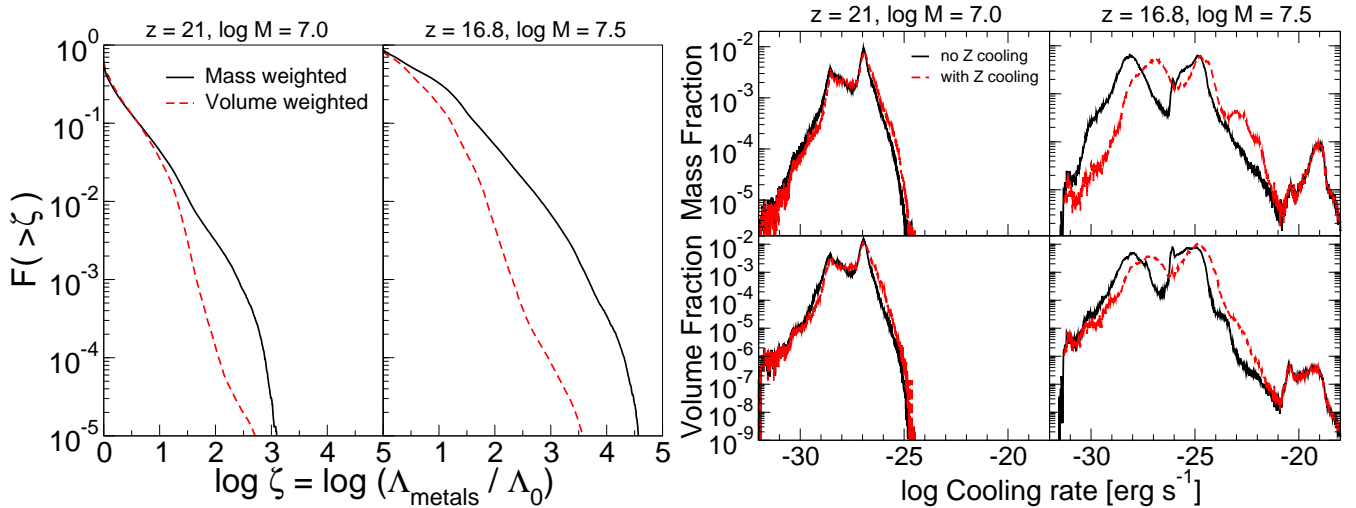}{f14b_color}
\caption{\label{fig:cool_ratios} Left: Fraction of mass (black solid)
  and volume (red dashed) of the sphere with a radius of 1 kpc,
  centered on the most massive halo at $z$ = 21 (left) and 16.8
  (right), that would have a high cooling rate by a factor of $\zeta$
  if we would consider metal-line cooling.  Right: Mass-weighted (top)
  and volume-weighted (bottom) histograms of cooling rates of the same
  volume at redshift 21 (left) and 16.8 (right).  The solid black
  lines are the cooling rates with primordial abundances, whereas the
  red dashed lines consider metal-line cooling.  The increased cooling
  at 10$^{-27}$ erg s$^{-1}$ is caused by metal-enriched cool gas, and
  the increased cooling above $10^{-25}$ originates from warm
  ($\gsim10^4$ K) gas, mainly in SN remnants.}
\end{center}
\end{figure*}
%%%%%%%%%%%%%%%%%%%%%%%%%%%%%%%%%%%%%%%%%%%%%%%%%%%%%%%%%%%%%%%%%%%%%%%%

When the most massive halo reaches a virial temperature of $10^4$ K at
$z$ = 16.8, metals are thoroughly mixed in the halo, and its mean
metallicity is \zstar{-2.6}, where \feighty~is in units of 80 \Ms~of
metal ejecta per PISN.  Turbulence created from the dynamic assembly
of this \tvir~= 10$^4$ K halo that involves outflows generated by
feedback, cold accretion through filaments, halo mergers, and
virialization (Paper I) appears to thoroughly mix the metals
\citep{Pan07}.  As stars form in this halo, the metallicity of this
halo fluctuates around this value because stars continue forming but
ejecting most of their metals into the IGM.  Also the filaments are
still mostly pristine and provide a source of nearly metal-free cold
gas.  The volume averaged metallicity of the enriched IGM ($\delta <
10$) is \zstar{-3.0}, compared to the filaments and halos ($\delta >
10$) that are less enriched with \zstar{-3.6} \citep[cf.][]{Cen08}.
Considering the total mass of heavy elements, there are 360 \Ms~of
heavy elements in the IGM, about 40\% of the total ejecta from all
PISNe.  The metal volume filling fraction is 3.5\% of the volume where
we allow star formation to occur.  This percentage should be higher
than the cosmic mean at this redshift because this comoving volume of
(205 kpc)$^3$ is biased with an overdensity $\delta \equiv
\rho/\bar{\rho}$ = 1.8.  Thus star formation rates are greater than
the mean since there are more high-$\sigma$ peaks, and the metal
filling fraction should scale with this bias.

The $\rho$-T phase diagram in Figure \ref{fig:hotphase} is colored by
the mean metallicity of the gas.  These data are taken immediately
before the formation of a star with a metallicity of \zstar{-3.6}.
There are three distinct metallicity states in the halo.  The majority
of the gas in the halo has a density between 10$^{-3}$ and 1\cubecm.
This gas has a mean metallicity of \zstar{-2.5}.  At higher densities,
the metallicity is slightly lower at \zstar{-3.5}.  The same
preferential enrichment of diffuse regions may have caused the lower
metallicities in this dense cloud.  The third phase is the warm, low
density ($\rho < 10^{-4}\cubecm$) gas that exists in recent SN
remnants and has solar metallicities and greater.  

The hot phase produced by a SN is super-solar.  The high-density tail
of the ejecta is the SN shell and is mixing with the lower metallicity
ambient medium.  As the ejecta expands and cools, it will contribute
to the warm, low density, solar metallicity material, whose cooling
time is greater than the Hubble time.

Although we do not consider metal-line cooling in the enriched
material, it is useful to determine the importance of metal-line
cooling in early galaxies.  As stated above, heavy elements within the
most massive halo are well-mixed by turbulence and should contribute
to the cooling of the gas as it is enriched above $\sim$$10^{-5}
Z_\odot$ \citep[e.g.][]{Maio07, Smith08}.  We calculate cooling rates
with contributions from metal-line cooling in the most massive halo in
SimB-SN.  We stress that the following cooling rates from metals are
upper limits since every Population III star produces a PISN.  For gas
with temperatures below $10^4$ K, we calculate the additional cooling
from carbon, oxygen, and silicon using the rates and reactions from
\citet{Glover07}.  For hotter gas, we use the metal cooling rates from
\citet{Sutherland93}, i.e. $\Delta\Lambda(Z) = \Lambda_{\rm{net}}(Z) -
\Lambda_{\rm{net}}(Z=0)$, where $\Lambda_{\rm{net}}$ is the net
cooling rate listed in \citeauthor{Sutherland93} and $Z$ is the
metallicity relative to solar.  Figure \ref{fig:metalcool} compares
the cooling rates with only primordial abundances
\citep[i.e.][]{Abel97} and with metal-line cooling in a
density-temperature phase plot.  We plot the data from a sphere of 1
proper kpc, centered on the most massive halo when it forms its second
star ($z = 21$) and the final redshift ($z = 16.8$).  The pixel
intensity represents the ratio $\zeta = \Lambda_{\rm{metals}} /
\Lambda_0$, the cooling rate with metal-line cooling and with
primordial abundances.  The gas that experiences the greatest metal
cooling compared to cooling from H, He, and \hh~is either in old SN
remnants and \ion{H}{2} regions that have $T \gsim 10^4$ K or in cool
($T < 1000$ K) and diffuse ($\rho \sim 1 \cubecm$) gas where
fine-structure metal lines start to dominate over \hh~cooling.  The
warm ($T \approx 10^4$ K) ISM cools by atomic hydrogen line cooling
and metal line cooling and the recent SN remnant ($T > 10^5$ K) cools
mainly by Compton cooling.

We plot the cumulative sum of mass and volume that as a function of
$\zeta$ at $z$ = (21, 16.8) in the left panel of Figure
\ref{fig:cool_ratios}.  Approximately 12\% (41\%) of the mass in the
most massive halo at redshift 16.8 (21) have no significant metal-line
cooling.  At the final redshift, over a quarter of the gas mass has
cooling rates increased over an order of magnitude.  However $z = 21$,
this mass fraction drops to 4\%.  In the right panel, we also plot the
probability density function of cooling rates in both the enriched and
primordial cases.  We see the features from \ion{H}{2} regions, SN
remnants, and cool gas in Figure \ref{fig:metalcool} also in Figure
\ref{fig:cool_ratios}.  The shift of the first bump that was
originally centered at $10^{-28}$ erg s$^{-1}$ was caused by the cool
metal-enriched gas, and the shift at higher rates occurs in \ion{H}{2}
regions.  Taken as an upper limit of metal enrichment, it appears that
metal cooling considerably alters the cooling characteristics of the
gas within these early dwarf galaxies and should be accounted for in
future simulations that study the early dwarf galaxy formation or the
transition to Pop II star formation.

\begin{figure}[t]
  \begin{center}
    \epsscale{1.15}
    \plotone{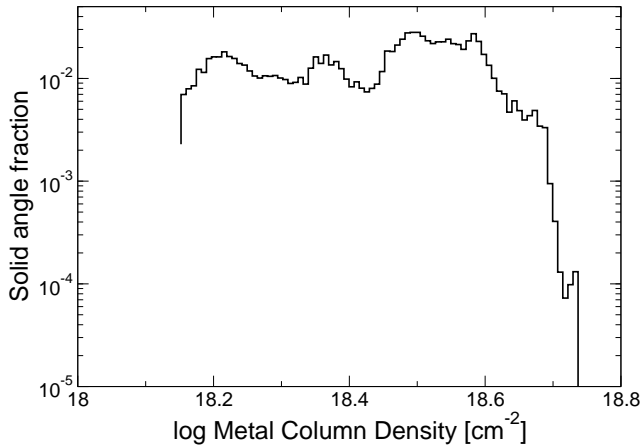}
    \caption{\label{fig:dust} Distribution of metal column densities
      between the center of the most massive halo and a sphere of
      radius 3 proper kpc at $z = 16.8$ in SimB-SN.}
  \end{center}
\end{figure}

As the metallicities in early dwarf galaxies increase, dust absorption
may become important.  The dust extinction cross-section in the Milky
Way is $\sigma_{d,MW} \simeq 3 \times 10^{-21}$ cm$^2$ per hydrogen
nucleus \citep{Hollenbach79, Cardelli89}.  If we assume the dust
extinction properties do not change at lower metallicities, the
cross-section scales with metallicity, giving $\sigma_d = (Z/Z_\odot)
\sigma_{d,MW}$.  The optical depth to dust absportion is simply
\begin{equation}
  \label{eqn:dust}
  \tau_d = \int n_H \sigma_d \; dl = \frac{\sigma_{d,MW}
    N_{\rm{metals}}}{Z_\odot},
\end{equation}
where $N_{\rm{metals}}$ is the column density of metals and $Z_\odot =
0.0204$.  In Figure \ref{fig:dust}, we calculate the distribution of
$N_{\rm{metals}}$ between the center of the most massive halo and a
sphere of radius 3 kpc with our ray tracer, exactly like in the
\hh~case in Figure \ref{fig:lw_tau}.  Column densities of metals are
distributed between $2 - 5 \times 10^{18}$~cm$^2$, corresponding to an
upper limit of $\tau_d = 0.3 - 0.7$.  Hence dust shielding should not
be significant in the formation of early dwarf galaxies.  More massive
halos, however, may be subject to dust shielding as the metallicity
and total hydrogen column density of the halo increase.

% Each panel in Figure \ref{fig:metalcool} weights the pixels by mass,
% average gas density, average temperature, and average metallicity.
% The peak of increased cooling at $\Lambda = 10^{-26}$ erg s$^{-1}$
% originates from recent, warn SN remnants.  All of the additional
% cooling below that rate comes from cool ($T < 10^4$ K) gas.
% Predictably, higher density gas with $Z/Z_\odot \approx 10^{-2}$
% experience the most increased cooling, in some cases up to a factor
% of $10^5$ greater.

%%%%%%%%%%%%%%%%%%%%%%%%%%%%%%%%%%%%%%%%%%%%%%%%%%%%%%%%%%%%%%%%%%%%%%%%
%
% METAL COLUMN DENSITY
%
%%%%%%%%%%%%%%%%%%%%%%%%%%%%%%%%%%%%%%%%%%%%%%%%%%%%%%%%%%%%%%%%%%%%%%%%
%%%%%%%%%%%%%%%%%%%%%%%%%%%%%%%%%%%%%%%%%%%%%%%%%%%%%%%%%%%%%%%%%%%%%%%%

\subsection{Metal-enriched Star Formation}

\begin{figure}[t]
\begin{center}
\epsscale{1.15}
\plotone{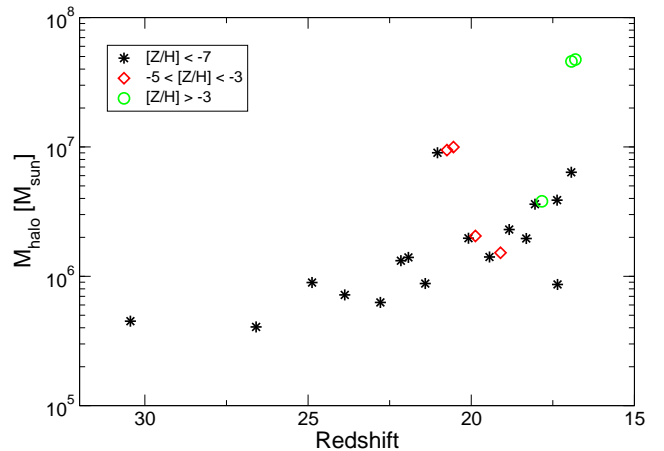}
\caption{\label{fig:SF} Star formation times of SimB-SN.  The
  $y$-axis shows the total mass of the host halo.  The symbols
  indicate the different metallicities of the stars and are labeled in
  the legend.}

%% The dashed line marks the mass of a halo with a virial temperature of
%% 4000 K, approximately at which multiple sites of star formation stars
%% to occur.

\end{center}
\end{figure}

The star formation times of SimB-SN is depicted in Figure
\ref{fig:SF} by plotting the total mass of the host halo versus the
formation redshift.  The different symbols represent the metallicity
of the star.  Before redshift 20, zero metallicity stars form in
low-mass halos, whose masses increase from $5 \times 10^5 \Ms$ to $2
\times 10^6 \Ms$ due negative feedback from photoevaporation of
low-mass halos.  The amount of photoevaporation in nearby halos is
apparent in the density projections in Figures
\ref{fig:proj_a}--\ref{fig:proj_b} where most nearby substructure is
lacking in the simulations with star formation.  This will further
suppress star formation already hindered by the LW radiation
background \citep{Machacek01}.

The first instance of a metal enriched ([Z/H] $>$ -6) star occurs at
$z = 20.7$ with a metallicity of \zstar{-4.0} in the most massive halo
that has a mass of $9.5 \times 10^6 \Ms$.  This star was triggered by
a SN blastwave, not in the IGM as envisaged by \citet{Ferrara98} but
within the same halo.  This SN explodes 470 kyr before the star and
provided the majority of heavy elements for the formation of this
enriched star.  The SN of the second star in the halo triggers yet
another round of star formation with a metallicity of \zstar{-3.8}
only 80 kyr afterwards.  The aggregate energy from these three SNe
expel most of the gas from the potential well.  The most massive halo
does not form any stars until $z = 16.9$ at which it undergoes two
episodes of star formation both with a metallicity of \zstar{-2.7}.

These metal-enriched stars could exceed the critical metallicity of
$\sim 10^{-4} \; Z_\odot$ \citep{Bromm01, Schneider06, Jappsen07b,
  Smith07}.  If this were to happen, the metal-enriched gas would
collapse and fragment into a stellar cluster \citep{Clark08}.  However
it is too preliminary to determine the differences in radiative and
chemical feedback between a Population III star and a Population II
stellar cluster.  As a point of reference, a $10^{-3} \; Z_\odot$ star
cluster with a Salpeter IMF would need to host 3500 \Ms~of stars to
produce equal amounts of hydrogen ionizing photons of what a 170
\Ms~Population III star produces \citep{Schaerer02, Schaerer03}.
Assuming a typical SN explosion energy of $7 \times 10^{48}$ erg
$M_\odot^{-1}$ in a stellar cluster with a Salpeter IMF, one would
similarily need 3900 \Ms~of stars to equal the explosion energy of a
170 \Ms~metal-free star.

%%%%%%%%%%%%%%%%%%%%%%%%%%%%%%%%%%%%%%%%%%%%%%%%%%%%%%%%%%%%%%%%%%%%%%%%
%
% STAR FORMATION HISTORY W/ METALLICITY
%
%%%%%%%%%%%%%%%%%%%%%%%%%%%%%%%%%%%%%%%%%%%%%%%%%%%%%%%%%%%%%%%%%%%%%%%%
%%%%%%%%%%%%%%%%%%%%%%%%%%%%%%%%%%%%%%%%%%%%%%%%%%%%%%%%%%%%%%%%%%%%%%%%

It should be noted that four stars in our calculation form with a
trace of metals ([Z/H] $\ll$ --6).  Three instances happen when a SN
blastwave overtakes a nearby halo.  Here the blastwave leaves the high
density material fairly pristine and the surrounding IGM is enriched
\citep{Cen08}.

In three halos with masses $1-4 \times 10^6 \Ms$, star formation is
triggered shortly ($<$ 3 Myr) after the death of a previous star in
the same halo.  The second star in the halo forms in a density
enhancement that is caused by an ionization front instability
\citep[e.g., see][]{Whalen08a, Whalen08b} when a SN blastwave
overtakes it.  The size of the overdensity is small enough ($\sim$30
pc) so the heavy elements can mix into the high density material
within a free-fall time.  In these three halos, this results in stars
with metallicities of [--5.3, --6.4, --4.3] $(M_{\rm{yield}} / \Ms)$
that form at $z$ = [19.9, 19.1, 17.8], respectively.  These
metallicities are uncertain due to the timing of star formation since
we neglect \hh~self-shielding.  If the density enhancement was
shielded from the LW radiation from the neighboring star and
\hh~cooling was still efficient, the star could have formed before the
SN blastwave enriched the cloud.  In that case, the resulting star
would have been metal-free.  Hence the metallicities of these induced
star forming regions should be considered with caution.  Studies with
a more accurate treatment of the \hh~self-shielding could better
capture the temporal sequence of metal enrichment and ionization front
instabilities.  Given the uncertainties in collapse times from our
simple star formation algorithm, masses, and feedback parameters, it
is improbable that a better \hh~line transfer calculation would lead
to significantly realistic results.

%% The gas that was reincorporated into the host halo was enriched by the
%% star that formed at $z = 31$.  The resulting SN again expels the gas
%% from the halo.  The gas is reincorporated again so stars can form in
%% the same halo around redshift 21.  Here three instances of star form
%% with metallicites of [--8.2, --8.1, --4.3] $(M_{\rm{yield}} / \Ms)$.

%% Similar instances of reincorporation of enriched material happens for
%% smaller halos at redshifts less than 20.  There is one metal enriched
%% star with a metallicity of \zstar{-4.8} that forms in a low-mass halo
%% with a mass of $2.7 \times 10^5 \Ms$ at $z = 17.2$.  This halo was a
%% satellite of a more massive star forming halo that was enriched by a
%% SN blastwave originating from the parent halo.

\section{DISCUSSION}
\label{sec:reionDiscuss}

We find that the combination of Pop III stellar feedback and continued
\hh~cooling in \tvir~$<$ 10$^4$ K halos alters the landscape of
high-redshift galaxy formation.  The most drastic changes are as
follows:

%\medskip

1. \textit{Dynamic assembly of dwarf galaxies}--- A striking
difference when we include Pop III radiative feedback are the outflows
and gas inhomogeneities in the halos and surrounding IGM.  The
outflows enrich the IGM and reduce the baryon fraction of the $10^4$ K
halo as low as 0.05, much lower than the cosmic fraction \Ob/\Om~=
0.17 \citep[cf.][]{Yoshida06a, Abel07}.  This substantially differs
from the current theories of galaxy formation where relaxed isothermal
gas halos hierarchically assemble a dwarf galaxy.  For instance in
simulation A, there are remarkable filamentary structures and a clumpy
ISM.  Furthermore, Pop III feedback increases the total baryonic
angular momentum of the system up to a factor of 3 without SNe and up
to 5 with SNe.

%\medskip

2. \textit{Pop III sphere of influence}--- Pop III feedback is mainly
a local phenomenon except its contribution to the UVB.  How far its
\ion{H}{2} region, outflows, and metal ejecta (if any) will
predominately determine the characteristics of the next generation of
stars.  Highly biased (clustered) regions are significantly affected
by Pop III feedback.  The first galaxies will form in these biased
regions and thus should be significantly influenced by its
progenitors.

%\medskip

3. \textit{Dependence on star forming progenitors}--- Although our
calculations with SNe only provided an upper limit of metal
enrichment, it is clear that the metallicity, therefore metal-line and
dust cooling and metal-enriched (Pop II) star formation, depends on
the nature of the progenitors of the dwarf galaxy.  If the galaxy was
assembled by smaller halos that hosted a Pop III star that did not
produce a SN, e.g., the galaxy would continue to have a top-heavy
initial mass function (IMF).

%\medskip

4. \textit{Complex protogalactic ISM}--- The interplay between stellar
and SNe feedback, cold inflows, and molecular cooling produce a truly
multi-phase ISM that is reminiscent of local galaxies.  The cool,
warm, and hot phases are interspersed throughout the dwarf galaxy,
whose temperatures and densities can span up to 6 orders of magnitude
at a given radius.

%\medskip

5. \textit{Metallicity floor}--- When the halo is massive enough to
host multiple sites of star formation, the metal ejecta does not
significantly increase the mean metallicity of the host halo.  There
seems to be a balance between (a) galactic outflows produced from SNe,
(b) inflowing metal-enriched gas, (c) inflowing pristine gas, and (d)
SNe ejecta that is not blown out of the system.  In our high yield
models, the metallicity interestingly fluctuates around $10^{-3}
Z_\odot$ in the most massive halo when this balance occurs at and
above mass scales $\sim$10$^7$\Ms.

%\medskip

Clearly the first and smallest galaxies are complex entities, contrary
to their low mass and generally assumed simplicity.  Our calculations
reflect the important role of Pop III stellar feedback in early galaxy
formation.  

These high-redshift galaxies have a $\sim$5--15\% chance of being
undisturbed by mergers until the present day, being ``fossils'' of
reionization \citep{Gnedin06}.  Dwarf spheroidals (dSph) galaxies are
some of the darkest galaxies in the universe, having high
mass-to-light ratios up to 100 \citep{Mateo98}.  Gas loss in dSph's
close to the Milky Way or M31 can be explained by gas tidal stripping
during orbital encounters \citep{Mayer07}.  However there are some
galaxies (e.g. Tucana, Cetus) removed from both the Milky Way and
Andromeda galaxies and cannot be explained by tidal stripping.  In
addition to ultraviolet heating from reionization \citep{Bullock00,
  Susa04} and intrinsic star formation \citep{MacLow99}, perhaps
stellar feedback from Pop III stars influenced the gas-poor nature of
dSph's.  Even at the onset of widespread star formation in the objects
studied here, the baryon fraction can be three times lower than the
cosmic mean, and the dwarf galaxy may never fully recover from the
early mass loss.  This initial deficit may play an important role in
future star formation within these low-mass galaxies and could
help explain the large mass to light ratio in isolated dSph's.

With the radiative and chemical feedback from the progenitors of the
early dwarf galaxies, we have an adequate set of cosmological
``initial conditions'' to study the transition from Pop III to Pop II
stars.  In this setup, the current metal tracer field can be used to
include metal line cooling.  Dust cooling may induce fragmentation of
solar mass fragments at metallicities as low as $\sim$10$^{-6}$ at
high densities \citep{Schneider06}.  However, metal-line cooling might
not be important at these low metallicities in diffuse gas.
\citet{Jappsen07a} showed that metal-line cooling at metallicities
below 10$^{-2} Z_\odot$ in low density gas does not significantly
affect the dynamics of a collapsing halo.  In a companion paper,
\citet{Jappsen07b} found that the fragmentation of a metal-poor ($Z =
10^{-3.5} Z_\odot$) collapsing object may depend more on the
conditions, e.g. turbulence and angular momentum distributions,
created during the assembly of such a halo than some critical
metallicity.

Perhaps when the protogalactic gas cloud starts to host multiple sites
of star formation, the associated SNe produce sufficient dust in order
for a transition to Pop II.  In lower mass halos, the SN ejecta is
blown out of the halo, and future star formation cannot occur until
additional gas is reincorporated into the halo.  However in these
halos with masses $\gsim$10$^7$\Ms, the SN does not totally disrupt
the halo.  A fraction of the SN ejecta is contained within halo and
could contribute to subsequent sites of star formation.  SN ejecta and
associated dust could instigate the birth of the first Pop II stars.

The metallicities of the first generation of metal-enriched stars
could depend on the metal mixing timescales in these early dwarf
galaxies.  Fortunately the dispersion of heavy elements from SNe into
the Galactic ISM is a rich field of study \citep[for a review,
  see][]{Scalo04}.  Metallicity dispersions in local stellar clusters
show fluctuations of 5--20\% around the mean, which suggest that the
ISM is well-mixed \citep[e.g.][]{Edvardsson93, Garnett00, Reddy03}.
In addition, very metal-poor ([Fe/H] $<$ --2.7) halo stars have very
little scatter in elemental abundances, suggesting that the dispersal
of the first metals that formed low mass stars originated from single
starbursts instead of single SN explosions \citep{Cayrel04}.  

In the ISM, laminar flows and turbulence driven by SN explosions
provides the impetus of metal mixing on large scales, which then
cascades to smaller and smaller length scales eventually reaching a
length scale associated with a Reynolds number $Re = 1$
\citep[e.g.][]{deAvillez02}.  In structure formation, turbulence can
also arise during the virialization of cosmological halos
\citep{Wise07a, Greif08}.  After the turbulent cascade creates
metallicity gradients on small enough scales, molecular diffusion will
homogenize the heavy elements.  \citeauthor{deAvillez02} performed AMR
simulations that include turbulent diffusivity to study the mixing
timescales in the ISM.  In their simulations and the ones presented
here, numerical diffusion provides the majority of the mixing at the
resolution limit.  They find that the mixing time is fairly
independent of the diffusion scale and could depend on some inertial
scale.  The origin and magnitude of diffusion does not significantly
affect the gas on large scales.  In their resolution study, mixing
timescales only decrease by 20\% when the resolution is increased by a
factor of 4.  Thus we could be overestimating the mixing timescales up
to a factor of a few because it is difficult to resolve the smallest
turbulent scale in cosmological simulations.  Furthermore, we expect
metals that are newly incorporated into the galaxy to be stirred by
the turbulence in assembling halos, which is sustained in
virialization and major mergers on the largest scales.

%\li Relation to local dSph metallicity gradients?

As discussed above, the metallicity of the most massive halo
fluctuates around $10^{-3} Z_\odot$.  This is intriguingly the same
value as a sharp cutoff in stellar metallicities in four local dSph's:
Sculptor, Sextans, Fornax, and Carina \citep{Tolstoy04, Helmi06}.
This differs with the galactic halo stars, whose metal-poor tail
extends to $Z/Z_\odot = 10^{-4}$ \citep{Beers05}.  We must take care
when comparing our results to observations since we made the
simplification that every Pop III star produces a PISN \citep[for a
detailed semi-analytic model of metal enrichment, see][]{Tumlinson06}.
As discussed in \citeauthor{Helmi06}, the galactic halo may be
composed of remnants of galaxies that formed from high-$\sigma$
density fluctuations, and dwarf galaxies originate from low-$\sigma$
peaks.  In this scenario, the objects (or its remnants) simulated here
would most likely reside in galactic halos at the present day.  If we
attempt to match this metallicity floor of $10^{-4} Z_\odot$ in the
galactic halo, this requires $\sim$8\Ms~of metals produced for every
Pop III star or roughly one in ten Pop III stars ending in a PISN.
More likely, the current nearby dSphs are hosted by larger dark matter
halos than we have been able to simulate to date.  Hence too simple
chemical evolution inspired extrapolations may be premature.

\section{SUMMARY}
\label{sec:summary}

Radiative feedback from Pop III stars play an important role in
shaping the first galaxies.  We studied the effects of this feedback
on the global nature of high-redshift dwarf galaxies, using a set of
five cosmology AMR simulations that accurately model radiative transfer
with adaptive ray tracing.  Additionally, we focused on the metal
enrichment of the star forming halos and their associated star
formation histories.  Our key findings in this paper are

%\medskip

1. Dynamical feedback from Pop III stars expel nearly all of the
baryons from low-mass host halos.  The baryon fractions in star
forming halos never fully recover even when it reaches a virial
temperature of 10$^4$ K.  The baryon fraction is reduced as low as
$\sim$0.05 with SNe feedback, three times lower than the cases without
stellar feedback.

2. Baryons on average gain angular momentum as they are expelled in
feedback driven outflows.  When it is reincorporated into the halo, it
increased the spin parameter up to a factor of 4 with radiative and
SNe feedback.

3. The accurate treatment of radiative transfer produces a complex,
multi-phase ISM that has densities and temperatures that can span up
to 6 orders of magnitude at a given radius.

4. Pair-instability SN preferentially enrich the IGM to a metallicity
an order of magnitude higher than the overdense filaments adjacent to
the sites of star formation.

5. Once a SN explosion cannot expel the gas in its host halo, the mean
metallicity fluctuates around \zstar{-2.6} as there may be a balance
between SN outflows, cold inflows, and contained SNe ejecta.

%\medskip

We conclude that Pop III stellar feedback plays an integral part to
early galaxy formation as it determines the characteristics of the
first galaxies.

\acknowledgments

This work was supported by NSF CAREER award AST-0239709 from the
National Science Foundation and partially supported in part by the
National Science Foundation under Grant No. PHY05-51164.  We thank
Marcelo Alvarez, Greg Bryan, and Naoki Yoshida for providing
constructive comments on an early draft.  Comments from an anonymous
referee enhanced this paper by suggesting that we explore the
importance of metal cooling, self-shielding, and metal mixing.  We are
grateful for the continuous support from the computational team at
SLAC.  We benefited the hospitality of KITP at UC Santa Barbara, where
this work was completed.  We performed these calculations on 16
processors of a SGI Altix 3700 Bx2 at KIPAC at Stanford University.

{}

\end{document}